%% file: main.tex
\documentclass[acmsmall]{acmart}

\usepackage{booktabs}
\usepackage{amsmath,amsfonts}
\usepackage[linesnumbered,ruled,vlined]{algorithm2e}
\usepackage{graphicx}
\usepackage{textcomp}
\usepackage{xcolor}
\usepackage{subfigure,url}
\usepackage{mathtools}
\usepackage{todonotes}
\usepackage{listings}
\usepackage{multirow}

\DeclarePairedDelimiter\floor{\lfloor}{\rfloor}

\newcommand{\rev}[1]{{{#1}}}
\newcommand{\revnew}[1]{{{#1}}}

\AtBeginDocument{%
  \providecommand\BibTeX{{%
    \normalfont B\kern-0.5em{\scshape i\kern-0.25em b}\kern-0.8em\TeX}}}

\setcopyright{acmcopyright}
\copyrightyear{2023}
\acmYear{2023}

\acmJournal{JACM}

\begin{document}

\title{Cache Optimization and Performance Modeling of Batched, Small, and Rectangular Matrix Multiplication on Intel, AMD, and Fujitsu Processors}

\author{Sameer Deshmukh}
\email{deshmukh.s.aa@m.titech.ac.jp}
\affiliation{%
  \institution{School of Computing, Tokyo Institute of Technology, AIST}
  \streetaddress{Ishikawadai Bldg. 9, 2-12-1 Ookayama, Meguro-ku}
  \city{Tokyo}
  \state{Tokyo}
  \country{Japan}
  \postcode{152-8550}
}

\author{Rio Yokota}
\email{rioyokota@gsic.titech.ac.jp}
\affiliation{%
  \institution{School of Computing, Tokyo Institute of Technology, AIST}
  \streetaddress{Ishikawadai Bldg. 9, 2-12-1 Ookayama, Meguro-ku}
  \city{Tokyo}
  \state{Tokyo}
  \country{Japan}
  \postcode{152-8550}
}

\author{George Bosilca}
\email{bosilca@icl.utk.co.us}
\affiliation{%
  \institution{Innovative Computing Laboratory, University of Tennessee at Knoxville}
  \streetaddress{Claxton 308C, 1122 Volunteer Blvd}
  \city{Knoxville}
  \country{USA}
  \postcode{37996}
}

\renewcommand{\shortauthors}{Deshmukh, Yokota and Bosilca.}

\begin{abstract}
Factorization and multiplication of dense matrices and tensors are critical, yet
extremely expensive pieces of the scientific toolbox. Careful use of low
rank approximation can drastically reduce the computation and memory
requirements of these operations.
In addition to a lower arithmetic complexity, such methods can, by
their structure, be designed to efficiently exploit modern hardware architectures.
The majority of existing work relies on batched BLAS libraries to handle the computation of
many small dense matrices.
We show that through careful analysis of the cache utilization, register accumulation using SIMD
registers and a redesign of the implementation,
one can achieve significantly higher throughput for these types of batched low-rank matrices 
across a
large range of block and batch sizes. We test our algorithm on 3 CPUs using diverse ISAs -- the 
Fujitsu A64FX using ARM SVE, the Intel Xeon
6148 using AVX-512 and AMD EPYC 7502 using AVX-2, and show that our new batching methodology is 
able to obtain more than twice the throughput of vendor optimized libraries for all CPU
architectures and problem sizes.
\end{abstract}



\maketitle
\input{introduction}
\input{perf-optimization-techniques}
\input{batching-methodology}
\input{ecm-description}
\input{ecm-optimization}
\input{experiments}
\input{conclusion}

\begin{acks}
This work was supported by JSPS KAKENHI Grant Number JP18H03248, JP20K20624, JP21H03447. This work is conducted as a research activity of AIST - Tokyo Tech Real World Big-Data Computation Open Innovation Laboratory (RWBC-OIL). This work is supported by “Joint Usage/Research Center for Interdisciplinary Large-scale Information Infrastructures” in Japan (Project ID: jh210024-NAHI).

Many thanks to colleagues from Tokyo Institute of Technology, 
AIST, RIKEN-CCS and University of
Tennessee at Knoxville who provided their valuable feedback 
for the improvement of this manuscript. Special thanks to 
Mohamed Wahib, 
Jens Domke and Chen Peng
for their valuable feedback on performance models and
optimization
of Fujitsu A64FX benchmarks.
\end{acks}

\bibliographystyle{ACM-Reference-Format}
\bibliography{refs.bib, references.bib, rioyokotalab.bib}

\end{document}

%% file: introduction.tex
\section{Introduction}
\label{sec:introduction}

Large dense matrices and tensors appear in applications such as Boundary integral methods,
machine learning, computational finance and multivariate regression. While direct
multiplication and factorization of such data sets is computationally expensive,
careful use of low rank approximation techniques can drastically reduce the compute
and memory cost of these methods with a controllable decrease in accuracy.
Examples of such applications are the use of hierarchical matrices~\cite{hackbusch2015} for
factorization of large dense matrices, tensor decomposition for multilinear systems~\cite{brazell2011} and
FMM in Deep Learning~\cite{nguyen_fmmformer_2021}.

Maximizing the computational efficiency of such methods has its unique challenges, different from both dense matrices and sparse matrices.
On one hand, it shares some of the challenges of sparse matrices, where the amount of arithmetic operations (Flops) per loaded data (Bytes) decreases with the rank of the low-rank blocks.
On the other hand, it is different from sparse matrices in the sense that the
low-rank blocks are not completely
sparse but consist of small dense matrices, providing some opportunities for data reuse and prefetch.
The resulting fine-grain regularity allows these methods to more efficiently
utilize SIMD operations compared to the efficacy of sparse matrices.
Optimizing the throughput of these structured low-rank matrices, however, is not 
an area that has been investigated sufficiently in the literature. We have attempted to 
address this gap by manually writing 
our own matrix multiplication kernels for low-rank matrices that 
greatly outperform vendor optimized libraries in respect to this particular issue.

This work focuses on optimizing the inner kernel for this new type of problem, where the matrix is neither dense nor sparse.
As such, this work runs counter to other existing work that focus more on the parallel scalability of structured low-rank matrices.
Although~\cite{charara2018,cao2019,pei2019, rouet_distributed-memory_2015, yu_distributed_2019} report hierarchical matrix factorization 
on many hundreds of nodes, the computation shows sub-optimal resource utilization mainly attributed to
library routines that are not efficiently tuned for handling the memory bound kernels of such factorization routines.
We revisit this, and address the improvement of the efficiency
of the low rank kernels that form a core of the computation.

In this paper we propose a new technique for optimizing a central component of structured low-rank 
matrices, the low rank matrix multiplication.
Our technique performs batched computation of low rank matrices and shows, through a
careful utilization of the different levels of cache, more efficiency than vendor optimized math libraries such as MKL,
\rev{AMD-}BLIS and SSL-2 (Scientific Software Library) for a variety of block 
and batch sizes.
Thus, when compared to vendor libraries, our method shows stronger scaling for a shared memory execution.
Specifically, we make two contributions in this paper:
\begin{enumerate}
\item An improved algorithm for batched computation of low rank matrix multiplication, 
that can achieve more than $2x$ greater throughput than vendor optimized libraries 
for all the CPU architectures and problem sizes tested;
\item Techniques for optimization and performance validation of low level kernels with extensive use of the
ECM~\cite{hofmann2020} (Execution-Cache-Memory)
performance model.
\end{enumerate}

The rest of this paper is organized as follows. In
Section~\ref{sec:low-rank-matrix-approximation} we review 
the existing literature
on low rank matrices and concretely define the low rank multiplication
operation optimized in this paper. In
Section~\ref{sec:perf-optimization-on-multiprocessors} we review the current
state of the art in obtaining optimal chip performance, along
with various performance modeling methodologies for guiding the development
of high performance implementations.
Section~\ref{sec:batching-methodology} describes the new
algorithm and proposed cache blocking methodology. Section~\ref{sec:ecm-perf-model} 
reviews the applicability of the ECM performance model for our problem on
the tested CPUs and demonstrates that use of the ECM model plays a pivotal
role in achieving the best possible performance from a given CPU. 
Section~\ref{sec:single-threaded-ecm-model} contains a detailed analysis of 
each kernel within our algorithm and presents the optimization on each of our 
target CPUs using the ECM performance model. We then provide the results
of our method for our target CPUs compared to various vendor-optimized
libraries in Section~\ref{sec:exp-evaluation} and finally conclude the
paper in Section~\ref{sec:conclusion}.

\section{Low Rank Multiplication}
\label{sec:low-rank-matrix-approximation}

The low rank approximation of a dense matrix allows capturing the most significant row
and column bases of the dense matrix. If the singular values
of the dense matrix (typically obtained from the Singular Value
Decomposition of the matrix) reduce very rapidly, we only need to retain the first few
singular values and associated basis vectors,
thus expressing the dense matrix with significant data compression. The number of
significant bases retained is the \textit{numerical rank} of the matrix.

\begin{figure}[tbp]
  \centering
  \includegraphics[width=\linewidth]{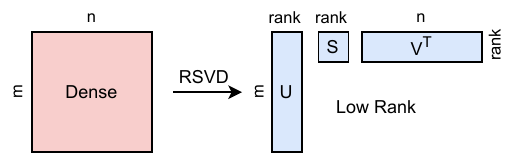}
  \caption{Representation of a low rank matrix using an
  algebraic method such as the Randomized Singular Value
  Decomposition (RSVD).}
  \label{fig:repr-low-rank-matrix}
\end{figure}

Low rank matrices are generated from the corresponding dense matrix block using
a decomposition like randomized SVD (Singular Value Decomposition)~\cite{halko2009},
Interpolative Decomposition~\cite{halko2009}, and Adaptive
Cross Approximation (ACA)~\cite{rjasanow_adaptive_2002}
(which are more efficient than SVD).
The \rev{dimensions of the} low rank approximation of a dense matrix of dimension
$m \times n$ \rev{using a rank of $rank$} can be represented using a tuple of 3
elements $(m,n,rank)$.
A dense matrix $A_{m \times n}$ is represented as a product of three matrices as shown in Eq.
\ref{eq:dense-to-low-rank}, and represented in~Fig.\ref{fig:repr-low-rank-matrix}.
\begin{equation}
    A_{m \times n} \approx U_{m \times rank} \cdot S_{rank \times rank} \cdot V_{rank \times n}
    \label{eq:dense-to-low-rank}
\end{equation}
Thus the total storage
requirement of the matrix reduces to $m \times rank + rank \times rank + rank \times n$,
a value significantly smaller than $m \times n$ memory necessary for the dense matrix,
if $rank$ is much smaller than $m$ and $n$.

\rev{An important component in the approximation and solution of hierarchical matrices as
shown in Section~\ref{sec:introduction} is the low rank multiplication algorithm 
shown in Algorithm~\ref{alg:low-rank-mult}}.
This algorithm involves multiplication between two `skinny' matrices $A_{V^T}$ and $B_U$ and 
two `small' matrices
$A_X$ and $B_X$ of the form $A_X \times A_{V^T} \times B_U \times B_X$. Multiplication between
such matrices is particularly challenging as a
result of their small sizes, which makes the computation heavily memory bound.
The technique that we develop in this paper optimizes this specific step \rev{by batching
a large number of independent low rank matrices together for improved performance}.
The low rank multiplication forms the first step of the rounded 
addition \cite{bebendorfStabilizedRoundedAddition2007}
algorithm for addition of low rank matrices, and the low rank matrix-vector multiplication.

\begin{algorithm}[tpb]
  \SetAlgoLined
  \LinesNumbered
  \KwIn{LowRank A($A_U, A_X, A_{V^T}$) of $(m,k,rank_{A})$, LowRank B($B_U, B_X, B_V^T$) of $(k,n,rank_{B})$}
  \KwResult{$G_{XY}$ of size $rank \times rank$}

  $C_{temp} = A_{V^T} \cdot B_U$ \\
  $E_{temp} = A_X \cdot C_{temp}$ \\
  $G_{XY} = E_{temp} \cdot B_X$ \\

  \caption{Low Rank matrix multiplication.}
  \label{alg:low-rank-mult}
\end{algorithm}

%% file: perf-optimization-techniques.tex
\section{Performance optimization on multi-processors}
\label{sec:perf-optimization-on-multiprocessors}

\subsection{Software methodologies for optimal hardware utilization}
\label{sec:software-methods-optimal-utilization}

\begin{figure}[tbp]
  \centering
  \includegraphics[width=0.7\linewidth]{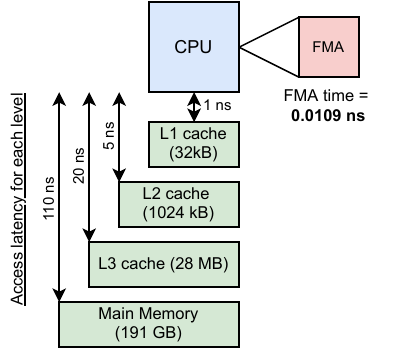}
  \caption{Access latency along the memory hierarchy when fetching data to execute an FMA on an Intel Xeon 6148 CPU (Skylake-X micro-architecture).}
  \label{fig:difference-xeon-6148}
\end{figure}

Having reached the limits of Moore's law, CPU architecture designers 
have moved from simply
improving CPU clock speed and die size to exposing multiple layers of parallelism
in their designs. Innovations such as SIMD architectures, Simultaneous Multi-threading (SMT) 
and ccNUMA designs have led to potential parallelism on every level of the CPU.
However, there is still a large difference in the time taken for arithmetic operations
vs. time taken for fetching them from memory. For example,
Fig.~\ref{fig:difference-xeon-6148} shows the time to execute an FMA operation on a single 
core on the Intel Xeon Gold 6148 CPU vs. the time taken to fetch a single double-precision
number from various layers of the cache hierarchy. Thus, the large ratio of memory vs. CPU speed (termed the
\rev{`}machine balance'~\cite{mccalpin1995}), has led software writers to adopt innovative
data locality optimizations in their designs in order to ensure that hardware spends most 
of its time on performing useful arithmetic calculations.

The dense linear algebra community has envisioned several approaches for optimizing
dense linear algebra routines using analytically modeled blocked 
algorithms~\cite{anderson1990, goto2008, vanzee2015, zee2016, yotov_is_2005,
low2016}, 
auto-tuning~\cite{whaley1998}, \revnew{systematic derivation of algorithms~\cite{gunnels2001}} and recursive cache oblivious algorithms~\cite{endo2020, frigo2012}.
Memory bound sparse matrix algorithms have similarly seen various 
innovations~\cite{im2004, vuduc2002, nishtala2007, alappat_ecm_2021}
where the implementation of register blocking and new sparse matrix formats have 
led to increased efficiency.
The tiny matrices arising out of low rank multiplication can be potentially batched
together for better SIMD and bandwidth utilization, as has been done by batched matrix multiplication routines from
MAGMA~\cite{haidar2015a}, Intel Math Kernel Library, and KokkosKernels~\cite{kim2017}.

Various approaches have been proposed for batching on both CPUs and 
GPUs~\cite{jiang2020, charara2019, abdelfattah2016, abdelfattah2016a, masliah2016}.
The LIBXSMM library \cite{heinecke2016, georganas2018} even optimizes batched matrix 
operations using register blocking and a JIT compiler. Alternate data layouts
that interleave data across batches have been shown to outperform simple batching in
some cases \cite{dongarra2017, kim2017}. This approach differs
from the other batching approaches in that it utilizes a different permutation of the
data across the batch dimension to keep the SIMD units busy. While efficient for
very small matrices, the packing and unpacking quickly becomes a bottleneck as the matrix
sizes increase. LibShalom~\cite{yang_libshalom_nodate-1} 
does away with the packing overhead for specific
matrix sizes and optimizes the instruction schedule to achieve high performance
of small matrix multiplications on ARM v8 CPUs. BLASFEO~\cite{frison2018, frison2020}
also provides batched LAPACK routines apart from matrix multiplication
and achieves better performance than vendor optimized libraries on various
CPUs. TSM2 and TSM2X~\cite{chen2019b, rivera2021} are specifically
built for optimizing tall-and-skinny matrix multiplication and
use code generation for tuning the usage of threads and the cache hierarchy
on the GPU.

\begin{figure}[tbp]
  \centering
  \subfigure[Small matrices] {
    \includegraphics[width=0.45\linewidth]{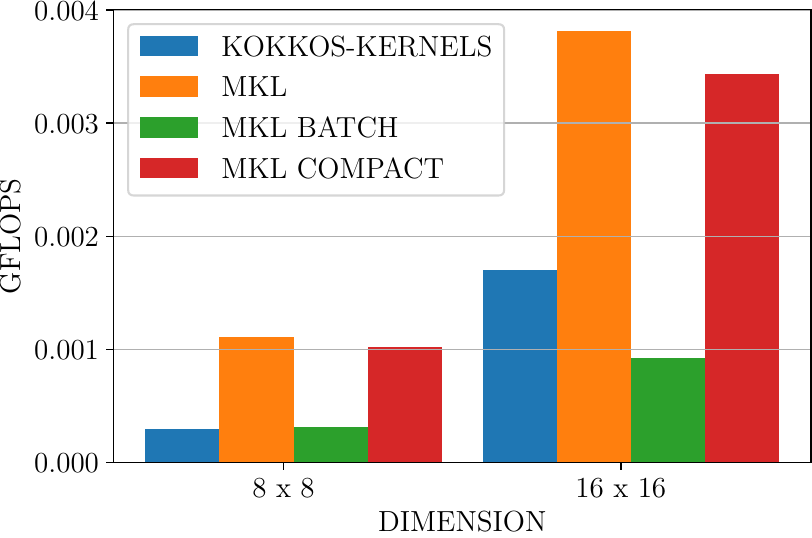}
    \label{fig:batched-small-matrix}
  }
  \subfigure[Skinny matrices] {
    \includegraphics[width=0.465\linewidth]{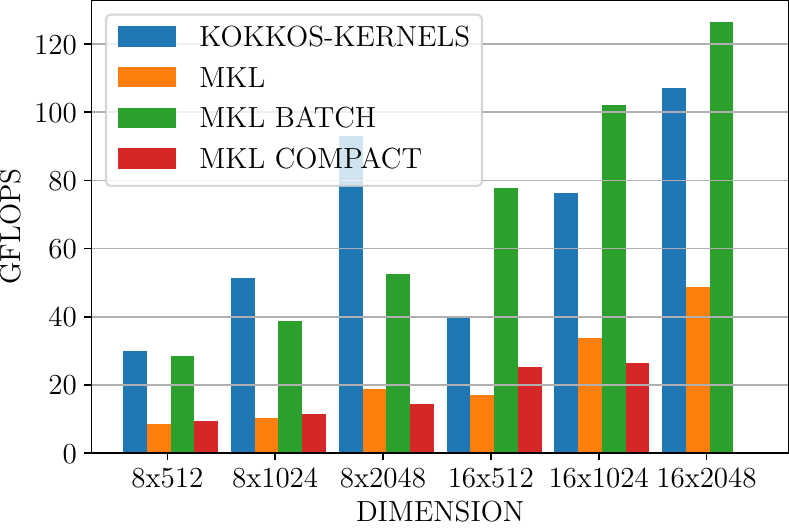}
    \label{fig:batched-skinny-matrix}
  }
  \caption{Performance of batched small and skinny matrices for a batch size of 10000 
  using 20 physical cores using various batching techniques on a Intel Xeon Gold 6148 CPU. 
  \rev{MKL uses non-batched MKL routines in a loop around the batch dimension. MKL BATCH performs
  batching with batched MKL routines without using memory interleaving across the batch dimension,
  and MKL COMPACT performs batching using memory interleaving as described by Dongarra et. al.~\cite{dongarra2017}. Kokkos-kernels \cite{kim2017} is a library
  designed from the ground up for obtaining efficiency with heterogeneous architectures
  and irregular matrix sizes.}}
  \label{fig:batched-skinny-small-utilization}
\end{figure}

Fig.~\ref{fig:batched-skinny-small-utilization} shows the performance
of various batching techniques \rev{(KOKKOS-KERNELS, MKL BATCH and MKL COMPACT) 
vs. not batched techniques (MKL)} 
for independent small matrices in Fig.~\ref{fig:batched-small-matrix} and for 
independent skinny matrices in  Fig.~\ref{fig:batched-skinny-matrix} 
\rev{for a constant batch size of 10,000.
Performing the same computation using batched
routines shows better performance in most cases. The batch size is fixed at 10,000 since it is found
to be a sufficiently large batch size to show the benefits of batched vs. non-batched routines.}
Clearly, while efficient implementations for batched skinny matrices exist,
small matrix operations are always under-performing. These operations are the least 
efficient part of many algorithms, 
and as such are the most promising candidates for enhancing the performance of 
low rank multiplication. \rev{Although Kokkos-kernels comes close to the performance
of vendor-optimized batched routines, it does so for only some specific cases and we 
therefore use only vendor-optimized libraries in our experimental evaluation in
Sec.~\ref{sec:exp-evaluation}.}

\subsection{Optimization with performance modeling}
\label{sec:comp-with-other}

Performance modeling is useful for predicting the ideal or peak performance of
an algorithm given architectural constraints. It can be used not only for
analytically deriving the peak performance, but also for indicating which area most benefits from direct optimization.

Techniques such as the roofline model \cite{williams2009} are useful
for gaining a measure of the memory-boundedness of an algorithm. However, these
`top-level' techniques do not dive deeper into the performance of 
the kernels in question. As a result, opportunities for optimization with the roofline model are 
difficult to identify.

\subsection{Performance modeling of LLC misses}

\begin{figure}[tbp]
  \centering
  \includegraphics[width=0.8\linewidth]{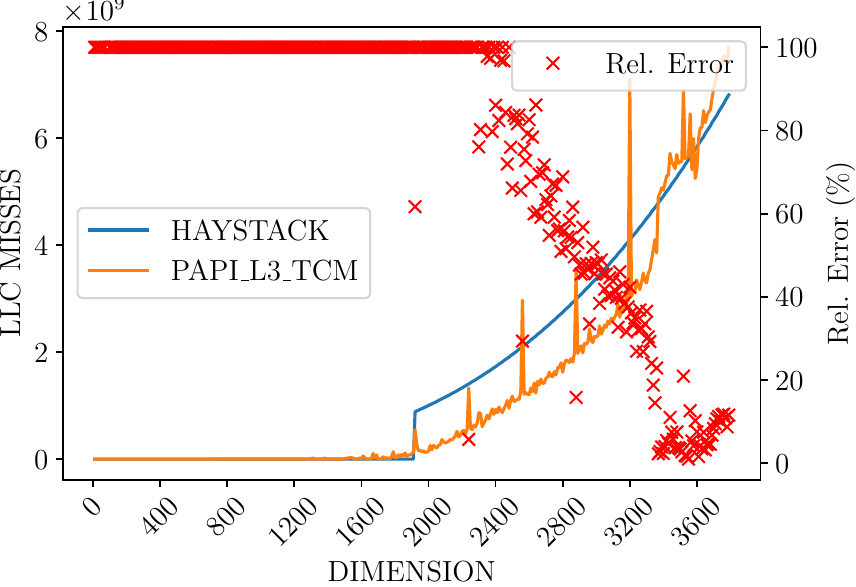}
  \caption{LLC misses of small matrix multiplication as predicted by haystack.
    The Y axis on the right shows the relative error between the calculated LLC misses by haystack
    vs. PAPI\_L3\_TCM measurements for $M = N = K$ as given in the X axis. Tests performed on a Intel
    Xeon Gold 6148 CPU (Skylake-X).}
  \label{fig:haystack-llc-prediction}
\end{figure}

A simple approach for modeling
the overall run time of memory bound applications is to measure the number of last-level-cache (LLC) misses~\cite{casas2014, huang2016, huang2016a, gysi2019}.
Several approaches ranging from analytical~\cite{huang2016a}
and semi-empirical modeling~\cite{hoefler2011} have been suggested
for this purpose. More advanced approaches involving simulation have been
suggested by~\cite{mattson1970, cavcaval2003} using stack
distances. This idea is extended by~\cite{gysi2019}, whose tool `haystack' allows
predicting the LLC misses using a faster simulation methodology than previously available.
However, as Fig.~\ref{fig:haystack-llc-prediction} shows, haystack is not
able to predict the LLC misses of smaller matrices.

%% file: batching-methodology.tex
\section{Batching Methodology}
\label{sec:batching-methodology}

\begin{algorithm}[tpb]
  \SetAlgoLined
  \LinesNumbered
  \SetNoFillComment
  \DontPrintSemicolon
  \KwIn{$A_{V^T}\_batch$, $B_U\_batch$, $A_X\_batch$, $B_X\_batch$}
  \KwResult{$G_{XY}\_batch$}
  \For(\tcc*[f]{Loop 1}){$batch\gets0$ \KwTo $BATCH\_SIZE$} {
    $A_{V^T} = A_{V^T}\_batch(batch)$ \\
    $B_U = B_U\_batch(batch)$ \\
    $A_X = A_X\_batch(batch)$ \\
    $B_X = B_X\_batch(batch)$ \\
    $G_{XY} = G_{XY}\_batch(batch)$ \\
    \For(\tcc*[f]{Loop 2}){$m \gets 0$ \KwTo $rank$} {
      \For(\tcc*[f]{Loop 3}){$n \gets 0$ \KwTo $rank$} {
        $C_{MN} = 0$ \\
        \For(\tcc*[f]{Loop 4}){$k \gets 0$ \KwTo $block\_size$} {
          $C_{MN} \mathrel{+}= A_{V^T}(m, k) \times B_U(k, n)$
        }
        \For(\tcc*[f]{Loop 5}){$x \gets 0$ \KwTo $rank$} {
          $E_{XN} = A_X(x, m) \times C_{MN}$ \\
          \For(\tcc*[f]{Loop 6}){$y \gets 0$ \KwTo $rank$} {
            $G_{XY}(x, y) \mathrel{+}=  E_{XN} \times B_X(n, y)$
          }
        }
      }
    }
  }
  \caption{Batched low rank multiplication expressed as 6 nested loops.}
  \label{alg:low-rank-mult-six-nest}
\end{algorithm}

\subsection{Looping order of Low rank multiplication}
\label{sec:looping-order-lr-mult}

The usual low rank multiplication is shown in Algorithm~\ref{alg:low-rank-mult}.
If done separately, this will lead to 3 nested loops
for each multiplication. Since there exist dependencies 
between these loops, we end
up writing the temporary blocks to memory after each step.
\rev{With a rewrite of the loops, we can perform the batched matrix multiplication 
using 6 nested loops as shown in Algorithm~\ref{alg:low-rank-mult-six-nest}}.
This way the result can be accumulated into a single variable $G_{XY}$ and written back 
into memory only once, at the end of the computation. We also notice that all temporary blocks used in this 
new algorithm are tiny matrices (size $rank \times rank$) and can fit within the vector registers
when $rank$ is sufficiently small.
We term the $A_X$ and $B_X$ as `small matrices' and $A_{V^T}$ and $B_U$ as `skinny matrices'.
\rev{Since the batch dimension is typically the largest dimension in a batched multiplication of
small sized matrices, performing the multiplication as specified  in Algorithm~\ref{alg:low-rank-mult-six-nest}
allows for more optimal utilization of bandwidth and therefore reduces the amount of time
spent in fetching data from main memory for a multi-threaded implementation. This has been experimentally
proven to be true in Sec.~\ref{sec:exp-evaluation}. Therefore, a combination of improved bandwidth utilization and
accumulation of intermediate results within SIMD registers allows our algorithm to achieve superior results
than vendor optimized libraries.}

\subsection{Locality optimization for low rank multiplication}
\label{sec:mult-threaded-low-latency}

\begin{algorithm}[tpb]
  \SetAlgoLined
  \LinesNumbered
  \KwIn{$A_{V^T}\_batch$, $B_U\_batch$, $A_X\_batch$, $B_X\_batch$}
  \SetKw{KwBy}{step}
  \KwResult{$G_{XY}\_batch$}
  \tcc{Pack as many small matrices in as possible in L3 cache with thread-level parallelism.}
  \For(\rev{\tcc*[f]{Loop 1A}}){$batch_{small} \gets 0$ \KwTo $BATCH\_SIZE$ \KwBy $B_{small}$} {
    $packed\_A_X \gets pack\_A_X\_matrices()$ \\
    $packed\_B_X \gets pack\_B_X\_matrices()$ \\
    \tcc{Pack $B_{skinny}$ skinny matrices into the L2 cache of each core.}
    \For(\rev{\tcc*[f]{Loop 1B}}){$batch_{skinny} \gets 0$ \KwTo $\frac{B_{small}}{B_{skinny}}$} {
      $offset \gets b_{small} \times B_{small} + b_{skinny} \times B_{skinny}$ \\
      $packed\_B_U \gets pack\_B_U()$ \\
      $packed\_A_{V^T} \gets pack\_A_{V^T}()$ \\
      \For(\rev{\tcc*[f]{Loop 1C}}){$batch \gets offset$ \KwTo $offset + B_{skinny}$} {
        $G_{XY} \gets G_{XY}\_batch(batch)$ \\
        \For(\rev{\tcc*[f]{Macro kernel. Loop 2}}){$m_c \gets 0$ \KwTo $m$ \KwBy $M_{PACK}$} {
          \For(\rev{\tcc*[f]{Loop 3}}){$n_c \gets 0$ \KwTo $n$ \KwBy $N_{PACK}$} {
            $C_{MN} = micro\_kernel\_cmn($ \\
            \Indp$m_c, n_c, packed\_B_U,$ \\
            $packed\_A_{V^T}$ \\
            \Indm$)$ \\
            \For(\rev{\tcc*[f]{Loop 5}}){$x_c \gets 0$ \KwTo $x$ \KwBy $X_{PACK}$} {
              $E_{XN} = micro\_kernel\_exn($ \\
              \Indp$m_c, x_c, packed\_A_X, C_{MN}$ \\
              \Indm$)$ \\
              \For(\rev{\tcc*[f]{Loop 6}}){$y_c \gets 0$ \KwTo $y$ \KwBy $Y_{PACK}$} {
                $G_{XY} = micro\_kernel\_gxy($ \\
                \Indp $packed\_B_X, E_{XN},$ \\
                $n_c, y_c$ \\
                \Indm$)$ \\
              }
            }
          }
        }
      }
    }
  }
  \caption{Batched low rank multiplication with a micro-kernel.}
  \label{alg:low-rank-mult-full}
\end{algorithm}

\begin{table}[tbp]
\begin{tabular}{|c|c|c|c|c|c|}
\hline
\textbf{CPU} & \textbf{$rank \ge VL$} & ${X_{PACK}}$ & \textbf{$Y_{PACK}$} & \textbf{$M_{PACK}$} & \textbf{$N_{PACK}$} \\ \hline
Fujitsu A64FX & YES/NO  & 8  & 8  & 8  & 8  \\ \hline
\multirow{2}{*}{\begin{tabular}[c]{@{}c@{}}Intel Xeon \\ Gold 6148\end{tabular}} & YES  & 4   & 16 & 8 & 16  \\ \cline{2-6} 
                                                                                 & NO  & 8  & 8  & 8 & 8  \\ \hline
AMD EPYC 7502                                                                    & YES/NO & 4 & 4 & 4 & 4 \\ \hline
\end{tabular}
\caption{\rev{Values of slicing variables according to architecture and $rank$. It is experimentally found that AMD
and Fujitsu micro kernels do not need modification irrespective of the $rank$ whereas the Intel micro kernels
perform best when the slice widths are changed if the rank is greater than the vector length ($VL$).}}
\label{tab:packing-values-cpu}
\end{table}
 
BLISLAB~\cite{note2016} separates the  implementation of the  dense matrix multiplication into a portable macro kernel written in a high level language such as C, and an architecture-specific micro kernel typically written using intrinsics or assembly code. This allows libraries \revnew{like BLIS} to be portable across a diverse set of machines and exposes thread-level parallelism~\cite{smith2014} in the macro kernel. This approach has been shown to attain near peak performance for a diverse set of CPU architectures as a result of enhanced data reuse.

We follow a similar approach for the low rank multiplication as shown in Algorithm~\ref{alg:low-rank-mult-full}. Assuming a three level cache hierarchy, loop 1A packs small matrices into the \revnew{last} level cache (L3) and loop 1B packs the skinny matrices into the L2 cache. Loop 1C then iterates over the batches of skinny matrices that are already packed in the cache. Thus loop 1 in Algorithm~\ref{alg:low-rank-mult-six-nest} is split into loop 1A, 1B and 1C in Algorithm~\ref{alg:low-rank-mult-full}. The data can then be streamed directly from the cache closest to the CPU (L1) when loading into registers. This can be changed to use only two cache levels in case of the A64FX CPU.

Algorithm~\ref{alg:low-rank-mult-six-nest} maintains portability across CPU architectures with varying cache sizes by changing the parameters $B_{small}$ and $B_{skinny}$ that control the number of small and skinny matrices being packed into cache, respectively. The number of small matrices $B_{small}$ being packed into the LLC is determined using Eq.~\ref{eq:1} (assuming {\bf double} as the type of our matrices). Eq.~\ref{eq:1} is obtained by dividing the number of bytes that the L3 cache can hold by the total number of bytes required for holding two $rank \times rank$ small matrices.

\begin{figure}[tbp]
  \centering
  \includegraphics[width=0.7\linewidth]{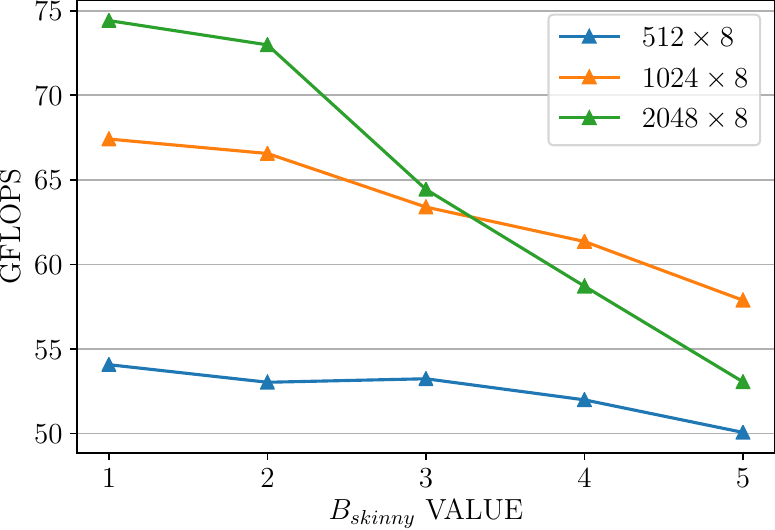}
  \caption{\rev{$B_{skinny}$ is the number of skinny matrices from each low rank
  matrix operand being packed into the L2 cache. This experiment shows the variation in performance as $B_{skinny}$
  is varied, keeping a sufficiently large batch size constant at $20000$ and using 20 physical cores of 
  an Intel Xeon Gold 6148 CPU. It can be seen that $B_{skinny} = 1$ leads to the best performance, 
  and this value is fixed in all future experiments.}}
  \label{fig:b-skinny-compare}
\end{figure}

The L2 cache typically has enough capacity to hold multiple skinny matrices from each operand of the low rank multiplication for a variety of block and rank sizes. Fig.~\ref{fig:b-skinny-compare} shows the effect of changing the number of skinny matrices from each low rank operand packed into the L2 cache. Experimentally, we find that packing only a single skinny matrix from each low rank operand leads to the best performance when using a sufficiently large batch size of 20,000 using an entire CPU socket (20 cores). Therefore, we fix $B_{skinny} = 1$ for all future experiments.

Algorithm~\ref{alg:low-rank-mult-full} shows how the loops shown in Algorithm~\ref{alg:low-rank-mult-six-nest} can be be expressed in terms of \textit{macro kernel} loops shown by the corresponding loops 2,3,5 and 6, and three micro kernels, each for accumulating the $C_{MN}$, $E_{XN}$ and $G_{XY}$ block. Each of these blocks is of size $S_{VEC} \times S_{VEC}$ where $S_{VEC}$ is the vector length of the CPU. Loop 4 from Algorithm~\ref{alg:low-rank-mult-six-nest} is absorbed into $micro\_kernel\_cmn()$ and optimized using assembly code. The macro kernel loops choose the slices of the packed blocks that must be computed by the micro kernels. The variables $M_{PACK}$, $N_{PACK}$, $X_{PACK}$ and $Y_{PACK}$ control the sizes of the slices that the macro kernel loops iterate over. These values are changed according to the architecture and the rank of the problem.

    For the case where $rank=S_{VEC}$, an entire matrix $G_{XY}$ of dimension $S_{VEC} \times S_{VEC}$ can 
    be computed within the
    registers without having to perform expensive reads and writes of the temporary
    matrices $C_{MN}$ and $E_{XN}$. In this case all the blocking variables in Algorithm~\ref{alg:low-rank-mult-full},
    i.e. $M_{PACK}$, $N_{PACK}$, $X_{PACK}$ and $Y_{PACK}$ will be equal to $rank$.
    In these instances the computation can be performed directly within the vector registers
    without a single write to memory. When the $rank$ is too large to fit into
    the SIMD registers, we perform blocking by using register blocks of different values
    so that the multiplication time of the skinny matrices is minimized and
    yields the best performance. Table \ref{tab:packing-values-cpu} shows the values of
    these variables when using a rank equal to the vector length and greater than the vector length for
    each CPU.

\begin{equation}
  \label{eq:1}
  B_{small} = \floor*{\frac{LLC_{bytes}}{2 \times rank \times rank \times sizeof(double)}}
\end{equation}

\subsection{Packing techniques for minimization of latency}

\rev{We tried out an alternate packing technique than that suggested in Sec.~\ref{sec:mult-threaded-low-latency}, 
where we packed the skinny matrices $B_U$ and $A_{V^T}$ matrices in the L3 cache and small matrices in
the L2 cache. However, this showed lesser performance than packing the small matrices in the
L3 cache. This is because the time for fetching small matrices into the L3 cache is minimized
when the outermost loop (loop 1A in Algorithm~\ref{alg:low-rank-mult-full}) performs this 
fetching since that loop has the most parallelism. Therefore,
the small matrices can be packed into the cache with maximum bandwidth utilization.}

Another technique is to pack the data across batches (similar to~\cite{kim2017}). If this is done, the 6 loop structure used in Algorithm~\ref{alg:low-rank-mult-six-nest} cannot be used since it relies on processing each low rank matrix one after another and not when the matrices are interleaved. \revnew{Moreover, the interleaved batched layout is slow for skinny matrices as shown in Fig.~\ref{fig:batched-skinny-small-utilization}.}

%% file: ecm-description.tex
\section{The ECM performance model}
\label{sec:ecm-perf-model}

The ECM (Execution-Cache-Memory) performance model \cite{wittmann2016} 
is an analytical technique for modeling the performance of steady-state loops. It 
models the ideal number of clock cycles necessary for execution of a single iteration
of a loop \rev{subject to the constraints of the algorithm and machine}.
This level of detail allows individual modeling of the kernels of an algorithm and 
cycles through an optimization process until they match or are close to the predicted performance.

Alappat et. al. have previously used the ECM model for optimizing sparse matrix vector multiplication
on the Fujitsu A64FX~\cite{alappat2020}. The ECM model has been used for optimizing conjugate gradient
based iterative solvers\cite{dreier2019, hofmann2020},
modeling the proportion of bandwidth utilized by overlapping kernels \cite{afzal2020}, and performance tuning and optimization of CFD applications \cite{wichmann2019, wittmann2016}.
Witmann et. al. \cite{wittmann2016} combine the ECM model
with an energy consumption model to optimize both the performance and power consumption
of a lattice-boltzmann CFD solver. They compare various performance modeling techniques 
and ultimately
utilize insights gained from the ECM model for gaining the most significant speedups
in their application.

The ideal number of clock cycles
for executing a single iteration of a loop using a single thread on a single physical core is
given by $T_{ECM}$. As shown in Eq. \ref{eq:ecm-basic-eq}, $T_{ECM}$ is composed of
various terms, which can be described as follows:
\begin{itemize}
    \item $T_{c}$ -- Clock cycles for pure compute instructions such as FMA and addition.
    \item $T_{L1_{L}}$ -- Clock cycles for loads from L1 cache into registers.
    \item $T_{L1_{S}}$ -- Clock cycles for stores from registers to L1 cache.
    \item $T_{l}$ -- Clock cycles for reads and writes between  $l-1$ and level $l$ cache.
    \item $T_{mem}$ -- Clock cycles for reads and writes between main memory and the \revnew{last level cache (LLC)}.
\end{itemize}

\begin{equation}
\label{eq:ecm-basic-eq}
T_{ECM} = max(T_{c}, f(T_{L1_{L}}, T_{L1_{S}}, T_{2} \dots T_{l}, T_{mem}))
\end{equation}

The main strength of the ECM model lies in the fact that it allows building an estimate of the overlap between levels of caches in a CPU. An important step in modeling the performance of any CPU using the ECM model involves first finding the function $f$ in Eq. \ref{eq:ecm-basic-eq} in order to quantify whether the reads and writes between the caches are simultaneous or in serial. This step differs between various CPU designs and has a non-trivial impact on the predicted performance.

Since the ECM model takes into account the individual instructions that comprise a loop, various bottlenecks such as assembly code generation by the compiler, inconsistent cache usage, and lack of Out of Order execution can be quickly pointed out.

We use the methodology provided by Hofmann et. al.~\cite{hofmann2020} for obtaining the ECM model for a given CPU architecture. \revnew{We first build the ECM model for the STREAM TRIAD kernel for each CPU. This is done by first building a model of the machine as shown in Sec. \ref{sec:ecm-machine-model}. We then build an application model specifically for STREAM TRIAD as shown in Sec. \ref{sec:ecm-application-model}. Finally, we can obtain the equation for $T_{ECM}$ for each CPU as shown in Sec. \ref{sec:ecm-overlap-hypothesis}. Since $T_{ECM}$ for each CPU (Table \ref{tab:machine-parameters-ecm}) remains constant for all steady-state loops for that CPU, the same equation can be used for our specific application of low rank multiplication.}

\subsection{Building the machine model}
\label{sec:ecm-machine-model}

\begin{table}[htbp]
  \centering
    \begin{tabular}{|c|c|c|c|}
      \cline{1-4}
      \multicolumn{1}{|l|}{\textbf{}}& \multicolumn{1}{|c|}{\textbf{AMD EPYC 7502}} & \multicolumn{1}{|c|}{\textbf{Fujitsu A64FX}} & \multicolumn{1}{|c|}{\textbf{Intel Xeon Gold 6148}}   \\ \cline{1-4}      Vector Length (bits)            & 512                                          & 512                                                 & 256                  \\ \cline{1-4}
      Instruction Set                & AVX2                                         & ARM SVE                                      & AVX512                                                \\ \cline{1-4}
      Microarchitecture              & Zen2                                         & ARM v8.2 SVE                                 & Skylake-X                                             \\ \cline{1-4}
      Cores                          & 32                                           & 48                                           & 20                                                    \\ \cline{1-4}
      FMA (/core)                    & 2                                            & 2                                            & 2                                                     \\ \cline{1-4}
      LOAD (/core)                   & 2                                            & 2                                            & 2                                                     \\ \cline{1-4}
      STORE (/core)                  & 1                                            & 1                                            & 1                                                     \\ \cline{1-4}
      Cache line size (bytes)        & 64                                           & 256                                          & 64                                                    \\ \cline{1-4}
      Cache write policy             & write-allocate                               & write-allocate                               & write-back                                            \\ \cline{1-4}
      Victim cache                   & Victim L3                                    & -                                            & Victim L3                                             \\ \cline{1-4}
      L1 load (bytes/cycle)          & 32                                           & 64                                           & 64                                                    \\ \cline{1-4}
      L1 store (bytes/cycle)         & 32                                           & 64                                           & 64                                                    \\ \cline{1-4}
      L2 load (bytes/cycle)          & 32                                           & 64                                           & 64                                                    \\ \cline{1-4}
      L2 store (bytes/cycle)         & 32                                           & 32                                           & 64                                                    \\ \cline{1-4}
      L3 load (bytes/cycle)          & 16                                           & -                                            & 14                                                    \\ \cline{1-4}
      L3 store (bytes/cycle)         & 16                                           & -                                            & 14                                                    \\ \cline{1-4}
      L1 size                        & $32 \times 32$ KiB                           & $48 \times 64$ KiB                           & $20 \times 32$ KiB                                    \\ \cline{1-4}
      L2 size                        & $32 \times 512$ KiB                          & $4 \times 8192$ KiB                          & $20 \times 1024$ KiB                                  \\ \cline{1-4}
      L3 size                        & $8 \times 16$ MiB                            & -                                            & $20 \times 1.375$ MiB                                 \\ \cline{1-4}
      Clock freq. (Hz)               & $2 \times 10^9$                              & $2 \times 10^9$                              & $2.2 \times 10^9$                                     \\ \cline{1-4}
    \end{tabular}%
  \caption{Various machine parameters used for building the ECM performance model for each CPU.}
  \label{tab:machine-parameters-ecm}
\end{table}

We first take into account various
machine parameters as shown in Table \ref{tab:machine-parameters-ecm} for our
target architectures. The A64FX is configured to run on \rev{`}normal' mode,
which means it runs at a frequency of $2 \times 10^9$ Hz.
For the Intel Xeon \rev{CPU} we disable Turbo Boost and 
assume the frequency that is obtained by running a simple Fused Multiply Add
loop using AVX-512 instructions. 
We chose these architectures since they use three diverse instruction
sets, AVX2, ARM SVE and AVX-512. This allows us to compare the
performance of the low rank matrix multiplication depending on various capabilities
provided by the Instruction Set Architecture (ISA), along with other machine parameters.
The cache \rev{LOAD/STORE bandwidths from various levels of cache} in 
Table \ref{tab:machine-parameters-ecm} can be determined by
formulating the STREAM usable bandwidth with empirical benchmarks and formulating 
and validating hypotheses about the bandwidth performance that fit the 
empirical measurements~\cite[Sec. 4]{hofmann2020}.


\subsection{Building the application model}
\label{sec:ecm-application-model}

The STREAM TRIAD \cite{mccalpin1995} kernel is a simple kernel of the form
$A(i) = B(i) + \alpha \times C(i)$. Assuming double precision, each
execution of the kernel requires loading 16 bytes of memory ($B(i)$ and $C(i)$)
and storing 8 bytes ($A(i)$) for a total of 24 bytes data transfer per iteration,
in addition to a single multiply and add operation that can be done as a single
operation using the fused multiply-add instruction. Note that the actual
amount of data transferred can vary according to the write policy of the
caches. Every memory access is assumed to be a full cache line 
transfer~\cite{wichmann2019}.

We build an ECM application model for the STREAM TRIAD kernel on similar lines
as has been done by  \cite{alappat2020, hofmann2020}.
The instructions that make up the STREAM TRIAD kernel, along with their latency 
and throughput on various architectures can be seen in 
Table~\ref{tab:tput-latency-stream-triad}.

\begin{table}[htbp]
    \centering
      \begin{tabular}{|l|l|l|l|l|l|l|}
        \cline{1-7}
        \textbf{}             & \multicolumn{2}{c|}{\textbf{AMD EPYC 7502}}                                                                & \multicolumn{2}{c|}{\textbf{Fujitsu A64FX}}                                                               & \multicolumn{2}{c|}{\textbf{Intel Xeon Gold 6148}}                                                          \\ \cline{1-7}
                              & \multicolumn{1}{c|}{Latency} & \multicolumn{1}{c|}{\begin{tabular}[c]{@{}c@{}}Reci.\\ TPut.\end{tabular}}  & \multicolumn{1}{c|}{Latency} & \multicolumn{1}{c|}{\begin{tabular}[c]{@{}c@{}}Reci.\\ TPut.\end{tabular}} & \multicolumn{1}{c|}{Latency} & \multicolumn{1}{c|}{\begin{tabular}[c]{@{}c@{}}Reci.\\ TPut.\end{tabular}}   \\ \cline{1-7}
        LOAD $A(i)$, REG0     &                            5 &                                                                     0.5     &                          11  &                                                                        0.5 &                           3  &                                                                      0.33    \\ \cline{1-7}
        LOAD $B(i)$, REG1     &                            5 &                                                                     0.5     &                          11  &                                                                        0.5 &                           3  &                                                                      0.33    \\ \cline{1-7}
        FMA REG0, alpha, REG1 &                            5 &                                                                     0.5     &                          9   &                                                                        0.5 &                           4  &                                                                      0.75    \\ \cline{1-7}
        STORE REG0, $C(i)$    &                            4 &                                                                     0.75    &                          9   &                                                                        1   &                           3  &                                                                      0.66    \\ \cline{1-7}
      \end{tabular}%
    \caption{STREAM TRIAD kernel with respective latencies and throughputs.}
    \label{tab:tput-latency-stream-triad}
\end{table}

Each instruction shown in Table \ref{tab:tput-latency-stream-triad} works in units
of one Vector Length (VL) corresponding to the length shown in Table \ref{tab:machine-parameters-ecm}.
The latency and throughput can be easily obtained by running identical
instructions in succession with dependencies between successive instructions
for latency and without dependencies for the throughput. Alternatively, the 
ibench~\cite{laukemann_automated_2018} tool can be used.

\subsection{Building the overlap hypothesis}
\label{sec:ecm-overlap-hypothesis}

\begin{figure}[htbp]
  \centering
  \subfigure[Fujitsu A64FX]{
    \label{fig:a64fx-stream-validate}
    \includegraphics[width=0.3\linewidth]{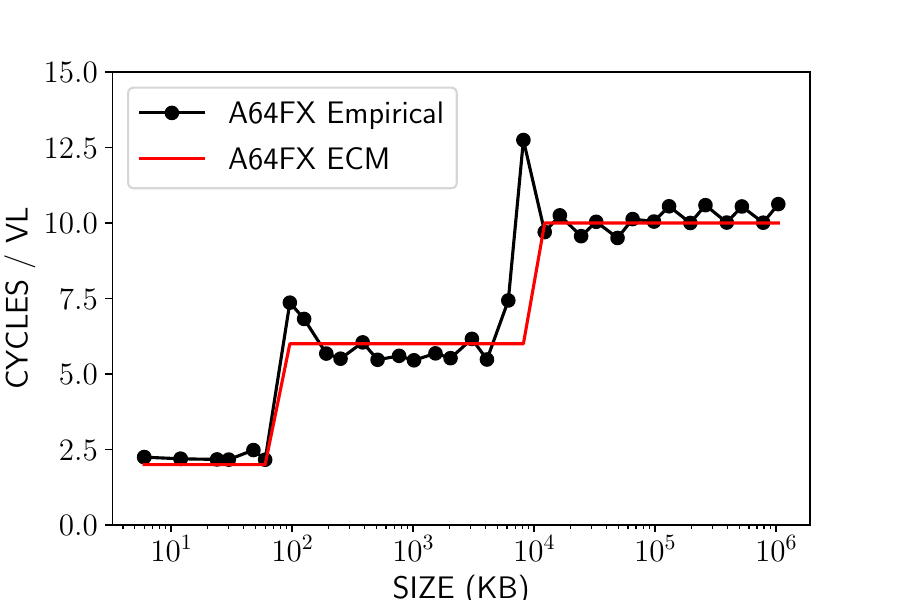}
  }
  \subfigure[Intel Xeon Gold 6148]{
    \label{fig:skx-stream-validate}
    \includegraphics[width=0.3\linewidth]{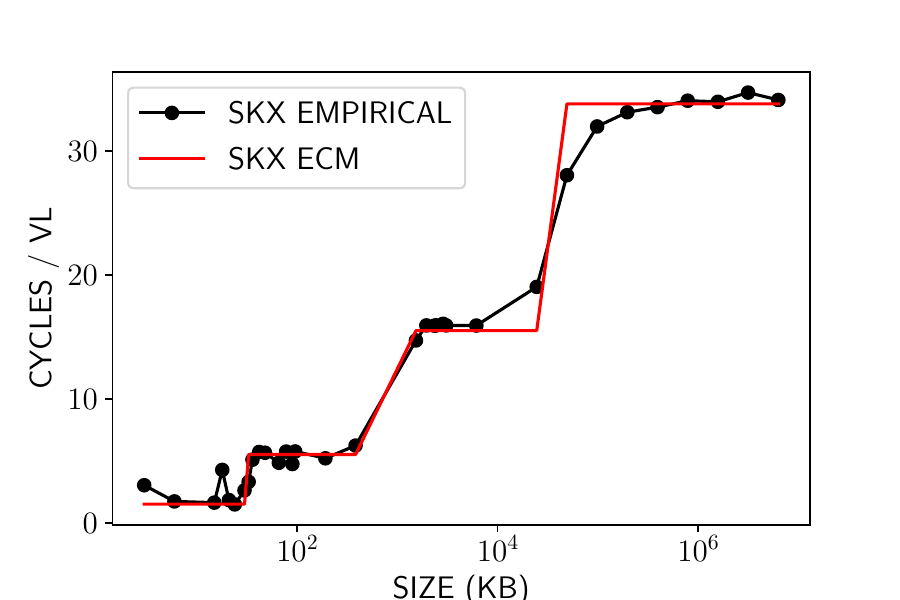}
  }
  \subfigure[AMD EPYC 7502]{
    \label{fig:epyc-stream-validate}
    \includegraphics[width=0.3\linewidth]{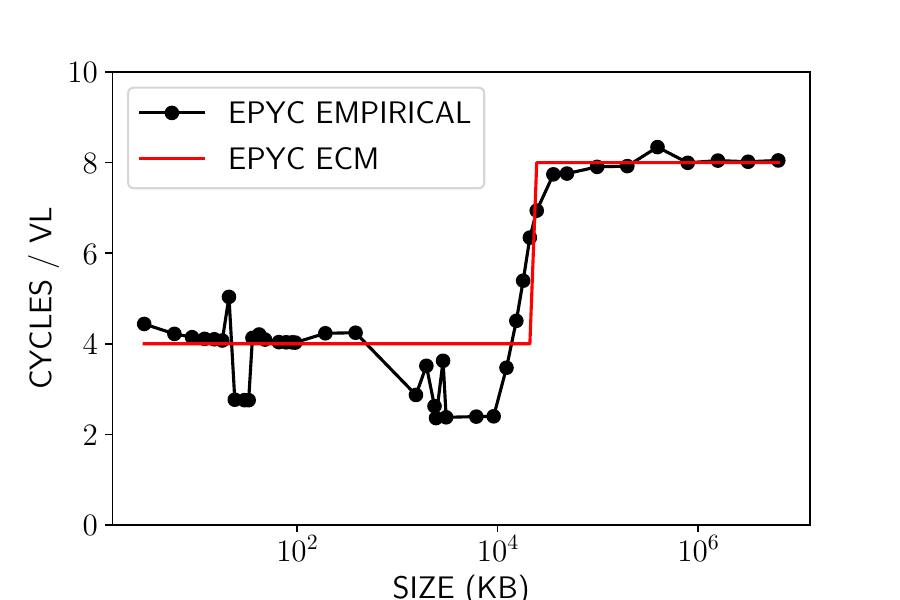}
  }
  \caption{Empirical vs. analytical clock cycles per \rev{vector length (VL)} for 
  each iteration of the STREAM TRIAD using a single core on each CPU. 
  \rev{As the size of the data increases along the X axis, the number of cycles required
  for fetching one VL goes up. The `steps' in each plot show how many cycles are
  needed when data fits into a particular level of cache. Each successive step shows
  that the data is being streamed from a progressively lower level of cache.
  The Fujitsu and Intel CPUs have a VL of 8 and the AMD CPU has a VL of 4.}}
  \label{fig:stream-triad-empirical}
\end{figure}

As shown by Hofmann et. al.~\cite{hofmann2020}, building the overlap hypothesis is an important step in construction of the ECM model for a given \rev{CPU}. Overlap hypotheses for the CPUs that we test have already been constructed for the Fujitsu A64FX~\cite{alappat2020}, and also for AMD and Intel \cite{hofmann2020, hofmann_ecm-based_2016} CPUs. In this section we  use the techniques shown by the aforementioned authors to derive our own assumptions about performance using the ECM model. We build ECM models for each \rev{CPU} as shown in Table~\ref{tab:cpu-ecm-relation}. The performance assumptions from these models are validated in Fig.~\ref{fig:stream-triad-empirical}.

\begin{table}[htpb]
  \begin{tabular}{|l|l|}
  \cline{1-2}
  \textbf{CPU} & \textbf{ECM Model} \\ \cline{1-2}
  Fujitsu A64FX        & $T_{ECM} = max(T_{c}, max(T_{L1_{L}} + max(T_{L1_{S}}, T_{L2}), T_{mem}))$   \\ \cline{1-2}
  Intel Xeon Gold 6148     & $T_{ECM} = max(T_{c}, T_{L1_{L}} + T_{L1_{S}} + T_{L2} + T_{L3} + T_{mem})$  \\ \cline{1-2}
  AMD EPYC 7502    & $T_{ECM} = max(T_{c}, T_{{L1}_{L}}, T_{{L1}_{S}}, T_{L2}, T_{L3}, T_{mem}) $ \\ \cline{1-2}
  \end{tabular}
  \caption{Derived ECM model assumptions for each CPU in our tests.}
  \label{tab:cpu-ecm-relation}
\end{table}

\rev{Fig.~\ref{fig:a64fx-stream-validate} shows the cycles per VL for the STREAM TRIAD
after disabling the compiler's aggressive prefetching and pipelining for the Fujitsu A64FX. 
Each step corresponds
to data being fetched from L1, L2 and main memory respectively. In practice, we observe
that the compiler does a lot of aggressive prefetching into the L2 cache and exhibits
behaviour as if the data were present in the L2 cache itself, which we use for constructing
the ECM model. Fig.~\ref{fig:epyc-stream-validate} shows that the AMD CPU goes from 4 cycles per 
VL to 8 after crossing the 16 MiB threshold as a result of the non-shared L3 cache present in
the chiplet-based Zen2 architecture, in spite of having a total 128 MiB of L3 cache.}

\rev{As per Table~\ref{tab:node-compile-config}}, even though the memory bandwidth of the 
\rev{AMD node is only 30\% higher than the Intel node, comparing the clock cycles needed to
fetch a single double precision digit from memory between Fig.~\ref{fig:skx-stream-validate} and
Fig.~\ref{fig:epyc-stream-validate} shows that the AMD CPU takes about half the clock cycles as
the Intel CPU} when the data size
exceeds that of LLC \rev{as shown in Table~\ref{tab:machine-parameters-ecm}}. This can be attributed to the
fact that the AMD CPU employs full memory overlap between all caches whereas the Intel CPU 
is completely non-overlapping, \rev{as can be seen in Table~\ref{tab:cpu-ecm-relation}}. 
Therefore, even though the Intel CPU employs the AVX-512 instruction
set with 512-bit long SIMD, the advantage still lies with the AVX2-enabled 256-bit
long AMD CPU as a result of fully overlapping communication between caches.

\revnew{Thus, Fig.~\ref{fig:stream-triad-empirical} shows that the Fujitsu A64FX takes the least number of clock cycles to fetch a single double precision number from main memory. This can be explained as a result of the better cache overlap design in the Fujitsu A64FX as shown in Table~\ref{tab:cpu-ecm-relation}.}


%% file: ecm-optimization.tex
\section{Single threaded optimization using the ECM performance model}
\label{sec:single-threaded-ecm-model}

The computation of the low rank multiplication kernels can be broadly divided
into kernel operations that involve packing the 4 matrices into caches, followed
by the \rev{computation in the $micro\_kernel\_cmn()$, $micro\_kernel\_exn()$ and 
$micro\_kernel\_gxy()$} as shown in Algorithm~\ref{alg:low-rank-mult-full}.
\rev{For brevity, we refer to $micro\_kernel\_cmn()$, $micro\_kernel\_exn()$ and
$micro\_kernel\_gxy()$ from Algorithm~\ref{alg:low-rank-mult-full} as the $C_{MN}$,
$E_{XN}$ and $G_{XY}$ kernels, respectively, after the intermediate products that they
compute.} 

\begin{figure}[tbp]
  \centering
  \includegraphics[width=.95\linewidth]{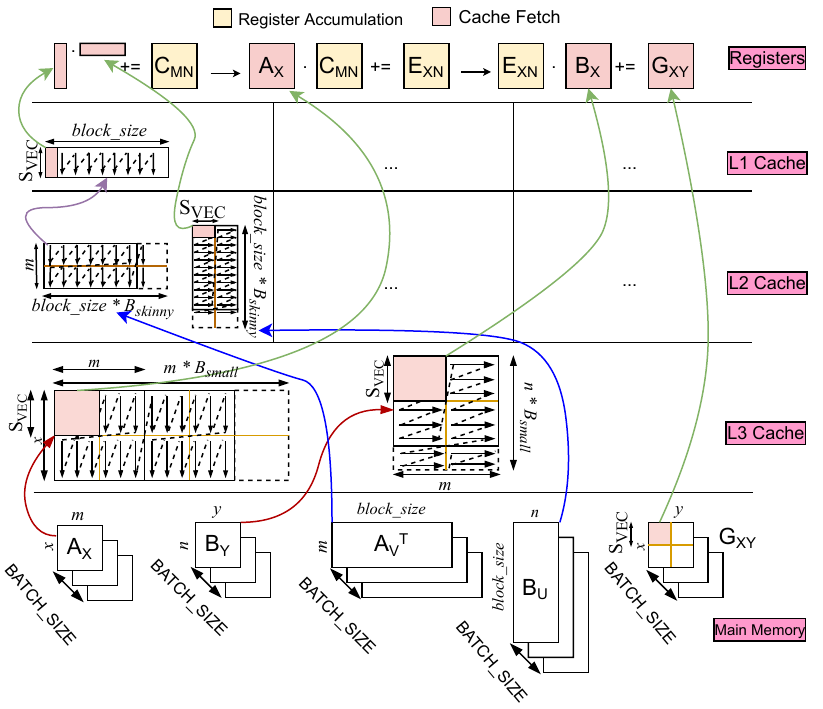}
  \caption{Diagram of the proposed low rank multiplication batching method.}
  \label{fig:low-rank-batching-diagram}
\end{figure}

We use the ECM model for optimizing the performance of the $C_{MN}$ kernel and the packing
of $A_{V^T}$ and $B_U$ since these are the most expensive parts. Using the ECM model
we have identified and addressed several bottlenecks in the default C++ code,
allowing us to reach very close to the maximum machine throughput. \revnew{Fig.~\ref{fig:low-rank-batching-diagram} shows the packing order of the operands into the caches. The small operands are packed into the shared L3 cache whereas the skinny matrices are packed into the per-core L2 and L1 caches.}
The $A_{V^T}$ is packed in column major whereas the $B_U$ is packed in row major
for facilitating loading into SIMD registers~\cite{low2016}.

\begin{table}[]
    \centering
    \resizebox{\textwidth}{!}{
      \begin{tabular}{llll}
        \cline{1-4}
        \multicolumn{1}{|c|}{\textbf{Instruction}} & \multicolumn{1}{l|}{\textbf{Description}} & \multicolumn{1}{c|}{\textbf{Latency}} & \multicolumn{1}{c|}{\textbf{\begin{tabular}[c]{@{}c@{}}Reci. \\ TPut.\end{tabular}}}   \\ \cline{1-4}
        \cline{1-4}
        \multicolumn{4}{|c|}{\textbf{AMD EPYC 7502 (Zen2)}} \\ \cline{1-4}
        \multicolumn{1}{|l|}{VMOVAPD(simple load)}   & \multicolumn{1}{l|}{Standard load with immediate addressing.}             & \multicolumn{1}{l|}{5} & \multicolumn{1}{l|}{0.5} \\ \cline{1-4}
        \multicolumn{1}{|l|}{VMOVAPD(simple store)}  & \multicolumn{1}{l|}{Standard store with immediate addressing.}            & \multicolumn{1}{l|}{4} & \multicolumn{1}{l|}{0.75} \\ \cline{1-4}
        \multicolumn{1}{|l|}{LEA}                    & \multicolumn{1}{l|}{Load effective address.}                              & \multicolumn{1}{l|}{2} & \multicolumn{1}{l|}{0.5} \\ \cline{1-4}
        \multicolumn{1}{|l|}{VGATHERQPD(gather)}     & \multicolumn{1}{l|}{Gather with stride of 8/16/32/512/1024/2048/4096 .}   & \multicolumn{1}{l|}{-} & \multicolumn{1}{l|}{6} \\ \cline{1-4}
        \multicolumn{1}{|l|}{VBROADCASTSD(simple)}   & \multicolumn{1}{l|}{Standard broadcast with a single double in memory.}   & \multicolumn{1}{l|}{4} & \multicolumn{1}{l|}{0.75} \\ \cline{1-4}
        \multicolumn{1}{|l|}{VFMADD231PD(simple)}    & \multicolumn{1}{l|}{FMA between 3 256 bit AVX registers.}                & \multicolumn{1}{l|}{5} & \multicolumn{1}{l|}{0.5} \\ \cline{1-4}
        
        \cline{1-4}
        \multicolumn{4}{|c|}{\textbf{Fujitsu A64FX (ARM v8.2 SVE)}} \\ \cline{1-4}
        \multicolumn{1}{|l|}{LD1D(simple)} & \multicolumn{1}{l|}{Standard load with immediate addressing.} & \multicolumn{1}{l|}{9} & \multicolumn{1}{l|}{0.5}  \\ \cline{1-4}
        \multicolumn{1}{|l|}{LD1D(gather, stride 8)} & \multicolumn{1}{l|}{Gather with stride of 8.} & \multicolumn{1}{l|}{-} & \multicolumn{1}{l|}{2}  \\ \cline{1-4}
        \multicolumn{1}{|l|}{LD1D(gather, stride 16/32/512/1024/4096)} & \multicolumn{1}{l|}{Gather with stride of 16/32/512/1024/4096.} & \multicolumn{1}{l|}{-} & \multicolumn{1}{l|}{4}   \\ \cline{1-4}
        \multicolumn{1}{|l|}{LD1D(gather, stride 2048)} & \multicolumn{1}{l|}{Gather with stride of 2048.} & \multicolumn{1}{l|}{-} & \multicolumn{1}{l|}{16}   \\ \cline{1-4}
        \multicolumn{1}{|l|}{ST1D(simple)} & \multicolumn{1}{l|}{Standard store with immediate addressing.} & \multicolumn{1}{l|}{9} & \multicolumn{1}{l|}{1}   \\ \cline{1-4}
        \multicolumn{1}{|l|}{LD1RD(simple)} & \multicolumn{1}{l|}{Standard broadcast with a single double in memory.} & \multicolumn{1}{l|}{9} & \multicolumn{1}{l|}{0.5}  \\ \cline{1-4}
        \multicolumn{1}{|l|}{FMLA(simple)} & \multicolumn{1}{l|}{FMA between 3 512 bit SVE registers.} & \multicolumn{1}{l|}{11} & \multicolumn{1}{l|}{0.5}  \\ \cline{1-4}
  
        \cline{1-4}
        \multicolumn{4}{|c|}{\textbf{Intel Xeon Gold 6148 (Skylake-X)}} \\ \cline{1-4}
        \multicolumn{1}{|l|}{VMOVAPD(simple load)} & \multicolumn{1}{l|}{Standard load with immediate addressing.}             & \multicolumn{1}{l|}{3} & \multicolumn{1}{l|}{0.33}  \\ \cline{1-4}
        \multicolumn{1}{|l|}{VMOVAPD(simple store)} & \multicolumn{1}{l|}{Standard store with immediate addressing.}           & \multicolumn{1}{l|}{4} & \multicolumn{1}{l|}{0.66} \\ \cline{1-4}
        \multicolumn{1}{|l|}{VGATHERQPD(gather)} & \multicolumn{1}{l|}{Gather with stride of 8/16/32/512/1024/2048/4096 .}     & \multicolumn{1}{l|}{-} & \multicolumn{1}{l|}{3} \\ \cline{1-4}
        \multicolumn{1}{|l|}{VBROADCASTSD(simple)} & \multicolumn{1}{l|}{Standard broadcast with a single double in memory.}   & \multicolumn{1}{l|}{1} & \multicolumn{1}{l|}{0.33} \\ \cline{1-4}
        \multicolumn{1}{|l|}{VFMADD231PD(simple)} & \multicolumn{1}{l|}{FMA between 3 512 bit AVX registers.}                 & \multicolumn{1}{l|}{5} & \multicolumn{1}{l|}{0.5} \\ \cline{1-4}

      \end{tabular}%
      }
    \caption{Latency and throughput of instructions depending on their operators for each CPU tested. All operations except memory-specific operations are performed on double precision floating
    point numbers.}
    \label{tab:cpu-instruction-table}
\end{table}
  
We calculate the latency and throughput of each instruction used in the kernels,
and report our findings in Table~\ref{tab:cpu-instruction-table}.
Note that we only consider bench-marking
the inner kernel execution and not the full computation,
therefore we ignore all instructions outside of the specified kernel. However, the full computation
is considered when reporting the final results in Sec. \ref{sec:exp-evaluation}.

We demonstrate the use of ECM modeling for the packing kernels when using ranks that are
multiples of the vector length of the machine.
Indeed, as highlighted in~\cite[3.4.2]{hofmann2020} the ECM prediction is \rev{a} function of 
the total number of full cache lines transferred, so we need to count in full cache 
lines even if only a part of the cache line will be used.
In case of strided cases with strides greater than $rank$, extra reads are considered as a
result of reading more cache lines in order to factor in the increased stride.

When reporting ECM model predictions, we report them as $T_{value} = (read) + (write)$ in order
to differentiate between read and write contributions in case of caches where both values
contribute to the outcome. In other cases the values show only read or only write contributions.

\subsection{Overall comparison of analytical and empirical kernel performance}
\label{sec:all-cpus-cmn-and-packing}

\begin{figure}[tbp]
  \centering
  \includegraphics[width=\linewidth]{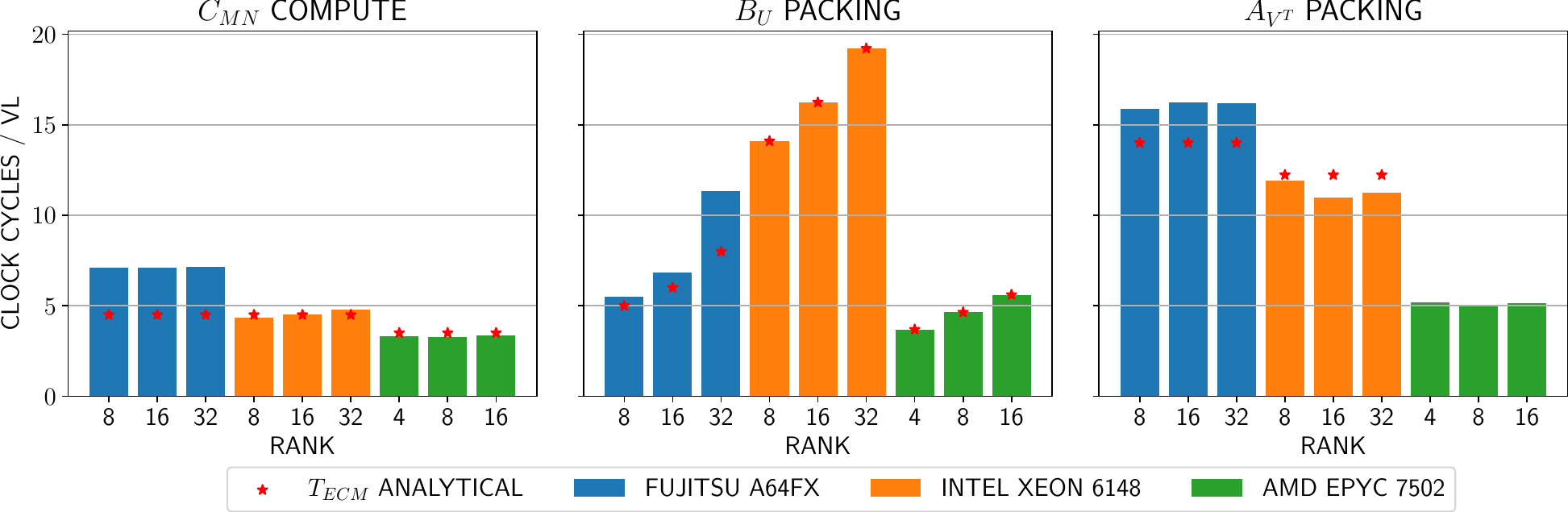}
  \caption{Comparison of analytical and empirical number of clock cycles for packing $A_{V^T}$
    and $B_U$ and computing $C_{MN}$. All tests are performed for a block size of $1024$ and batch size
    $10000$ using a single thread of execution. The ranks are varied as shown in the X-axis. The Y-axis
    shows the number of clock cycles taken per vector length.}
  \label{fig:overall-comparison-block-1024-compute-and-packing}
\end{figure}

We first show analytical vs. empirical results for the computation $C_{MN}$ and packing of
$B_{U}$ and $A_{V^T}$ for all CPUs. We address the challenges faced in reaching approximate peak ECM
performance for each CPU in subsequent sections. We then elaborate on the strategy used for 
determining $T_{ECM}$.

Fig.~\ref{fig:overall-comparison-block-1024-compute-and-packing} shows the empirical performance of the
three kernels vs. the analytical peak performance determined using the ECM model for each. It can be
seen that our code is able to reach theoretical peak performance as determined by the ECM model for
almost all cases. \rev{In subsequent experiments, we term the number of matrices in a batch as
the batch size and the longest dimension of the skinny matrices as the block size. We assume that both
low rank operands have equal rank and equal block size.} 

We report data only for block size 1024 as we do not observe a 
significant deviation in the results for larger block sizes, except for $A_{V^T}$
packing using block size 2048 on the Fujitsu
A64FX, which we elaborate on in Sec. \ref{sec:a64fx-packing-av-bu}.

\subsection{Optimization on the Fujitsu A64FX with the ECM model}
\label{sec:optimize-fugaku-ecm}

As shown in Table \ref{tab:cpu-instruction-table}, the throughput of the
\texttt{LD1D} instructions changes as the operands change.
The reciprocal throughput of this instruction when performing a gather operation
changes as the stride changes. We observe an anomaly when using a stride
of $8$ and $2048$, where the reciprocal throughput changes to 2 and 16 respectively.

\subsubsection{Packing $A_{V^T}$ and $B_U$ performance analysis}
\label{sec:a64fx-packing-av-bu}

We rely on manually inserting SVE gather and load intrinsics
for reducing the packing time, since we found that the Fujitsu compiler
does not make full use of vectorization. The $B_U$
matrix is packed in row major order, corresponding to the format in which it is already stored, so we can utilize {\tt LD1D (load)}
instructions for this purpose. Table \ref{tab:ecm-perf-strides-a64fx} shows
the ECM calculations for a variety of possible strides for $B_U$.

\begin{table}[tpb]
  \centering
    \begin{tabular}{|c|c|c|c|}
      \hline
      \multicolumn{1}{|c|}{\textbf{Variable}} & \multicolumn{1}{c|}{\textbf{Stride 8}} & \multicolumn{1}{c|}{\textbf{Stride 16}} & \multicolumn{1}{c|}{\textbf{Stride 32}} \\ \hline
      $T_{L1_L}$                                & 0.5                           & 0.5                                         & 0.5  \\
      $T_{L1_{S}}$                              & 1                                & 1                                          & 1 \\
      $T_{L2}$                                 & $(\frac{64}{64}) + (\frac{64 \times 2}{64}) = 3$ & $(\frac{64 \times 2}{64}) + (\frac{64 \times 2}{64}) = 4$  &
      $(\frac{64 \times 4}{64}) + (\frac{64 \times 2}{64}) = 6$  \\
      $T_{mem}$                                & $(\frac{64}{64}) + (\frac{64}{32} + \frac{64}{32}) = 5$ & $(\frac{(64 \times 2)}{64}) + (\frac{64}{32} + \frac{64}{32}) = 6$ &
      $(\frac{(64 \times 4)}{64}) + (\frac{64}{32} + \frac{64}{32}) = 8$  \\ \hline
      $T_{ECM}$                                & $5$                                & $6$                                           &  $8$ \\ \hline
    \end{tabular}%
  \caption{ECM performance breakdown for packing $B_U$ for various strides for Fujitsu A64FX.
    All measurements are reported in number of clock cycles.}
  \label{tab:ecm-perf-strides-a64fx}
\end{table}

The $A_{V^T}$ matrix is packed in column-major order, and similarly to the packing of
$B_U$ benefits from manually inserting SVE gather intrinsics.
%
While Fig. \ref{fig:overall-comparison-block-1024-compute-and-packing} shows the empirical vs.
analytical time taken when the block size is 1024.  For most strides, the $T_{ECM}$ can be calculated
by setting $T_{L1_{L}} = 4$, $T_{L1_{S}} = 1$,
$T_{L2} = (\frac{64 \times 8}{64}) + (\frac{64}{64} + \frac{64}{64}) = 10$
and $T_{mem} = (\frac{64 \times 8}{64}) + (\frac{64}{32} + \frac{64}{32}) = 12$. 
We observe however a performance anomaly for a stride of 2048 corresponding to the
increased latency of the {\tt LD1D} instruction shown in Table~\ref{tab:cpu-instruction-table}.
Table \ref{tab:packing-ecm-av-bu} shows the empirical vs. analytical performance when the block size
is 2048.

\begin{table}[]
  \centering
    \begin{tabular}{|c|c|c|c|c|c|}
      \hline
      \multicolumn{1}{|c|}{Block} & Rank & $A_V$ analytical & $A_V$ empirical \\ \hline
      2048                        & 8  & 26 & 26.32 \\
                                  & 16 & 26 & 26.22 \\
                                  & 32 & 26 & 26.30 \\ \hline
    \end{tabular}%
    \caption{Packing time per vector length for $A_{V^T}$ for Fujitsu A64FX for block size 2048.
      All measurements are shown in clock cycles per VL for a batch size of $10000$.}
  \label{tab:packing-ecm-av-bu}
\end{table}

\subsubsection{$C_{MN}$ kernel performance analysis}

The basic ARM SVE loop of the CMN kernel
uses one {\tt LD1D} instruction \rev{for loading 8 unique contiguous elements of the left operand}, 
8 FMA instructions \rev{for performing 8 SIMD multiplications} and 8 {\tt LD1RD} instructions
\rev{for loading 8 elements of the right operand duplicated within each register} for each
rank-1 update \rev{that results in a matrix of size $8 \times 8$.}

It can be seen that
there are 8 {\tt LD1RD} operations and 1 {\tt LD1D} operation
each taking 0.5 clock cycles, thus leading to a $T_{L1_{L}}$ of 4.5. There are a total
of 8 FMA operations, leading to a total $T_{c}$ of 4 clock cycles. Thus
$T_{ECM} = max(T_{c}, T_{L1_{L}}) = 4.5$ for a single rank-1 update.

We began by writing ARM ACLE instrinsics for the CMN kernel, but this turned
out to be \rev{taking many more clock cycles than the ideal shown by the ECM model} as 
a result of extra instructions generated in order to maintain portability between 
varying SVE lengths \cite[3.1]{stephens2017}.
We then enabled register length specific code generation and kept
a fixed vector length of $512$ using FCC compiler options, which led to 
more optimized code.

\rev{
Table~\ref{tab:cpu-instruction-table} shows that both the {\tt LD1RD} and {\tt FMA} instructions 
have a reciprocal throughput of $0.5$ cycles each. Given that one rank-1 
update generates an $8 \times 8$ matrix, a single {\tt LD1RD} and {\tt FMA} will together 
generate one row of this matrix, taking 1 clock cycle. 8 of these pairs will 
use 8 clock cycles. The reason why these instructions are executed likewise is
because we accumulate the intermediate products within the available SIMD registers,
which places an upper limit on the size of the rank-1 update that can be performed in the CMN block.
This, combined with the first {\tt LD1D} will take about 8.5 
clock cycles. As far as the {\tt LD1RD} and {\tt FMA}
instructions are concerned, exactly 64 bytes are computed for 64 bytes loaded, which
leads to a 1:1 ratio between the flops and bytes. For ideal {\tt FMA} throughput,
this ratio must be at least 2:1 so that two {\tt FMA} instructions can execute independently
on the two available FMA ports of the A64FX. }

\begin{figure}[htbp]
  \centering
  \includegraphics[width=\linewidth]{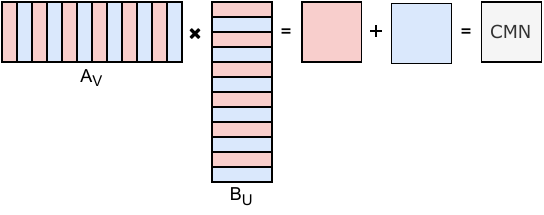}
  \caption{Rank-1 update for 8x8 matrix block using alternate rows and columns 
  of the skinny matrices and then adding the blocks at the end.}
  \label{fig:a64fx-rank-1-update-add}
\end{figure}

\rev{In order to overcome this limitation, we} utilize 8 extra registers 
and perform two separate rank-1 updates using alternate slices of 
the skinny matrices to improve the performance further. We then
add these slices at the end of the of the computation in order to obtain the
block in registers z0-7. This leads to the time dropping to about 7 cycles
per rank-1 update \rev{as a result of improved port pressure}. The technique can be described as in Fig.~\ref{fig:a64fx-rank-1-update-add}.

The ECM model assumes that instructions run at peak throughput, however that can never
be true in this case since the upper limit on the number of FMA instructions that can
be executed given the constraints on blocking has been reached. Therefore, about 7
cycles per rank-1 update is the best that can be achieved given the constraints on
the number of usable registers.

Table \ref{tab:fugaku-cmn-optimization} shows the comparison of the analytical $T_{ECM}$
compared to the performance obtained by successive optimizations. The usefulness of the
ECM model can be seen in support of optimization.

\begin{table}[]
  \centering
    \begin{tabular}{|c|c|c|c|c|c|}
      \hline
      \multicolumn{1}{|c|}{Block} & Rank & $T_{ECM}$ & SVE ACLE & SVE intrinsics & SVE multi-register intrinsics \\ \hline
      1024                        & 8    & 4.5      &    30.11       &  8.59        & 7.12 \\
                                  & 16   & 4.5       &    29.81       &  8.58        & 7.10 \\
                                  & 32   & 4.5      &    29.77       &  8.78        &  7.15 \\ \hline
    \end{tabular}%
  \caption{Comparison of successive optimizations to the CMN kernel compared to 
  $T_{ECM}$ for batch size of 10000 for the Fujitsu A64FX. All measurements are in 
  terms of clock cycles per rank-1 update.}
  \label{tab:fugaku-cmn-optimization}
\end{table}

\subsection{Optimization on Intel Xeon Gold 6148 with the ECM model}
\label{sec:skx-optimize-ecm}

\subsubsection{Packing $A_{V^T}$ and $B_U$ performance analysis}
\label{sec:skx-packing-ecm-model}

The ECM models for packing $B_U$ can be constructed by first noting the fact that packing one
VL (i.e. 8 doubles) requires one load (VMOVAPD(load)) and one store 
(VMOVAPD(store)) operation.
These operations have a throughput of $0.33$ and $0.66$, respectively. The breakdown
of the analytical ECM modeling is shown in Table \ref{tab:ecm-perf-strides-skx}.

\begin{table}[]
  \centering
    \begin{tabular}{|l|l|l|l|}
      \hline
      \multicolumn{1}{|c|}{\textbf{Variable}} & \multicolumn{1}{c|}{\textbf{Stride 8}} & \multicolumn{1}{c|}{\textbf{Stride 16}} & \multicolumn{1}{c|}{\textbf{Stride 32}} \\ \hline
      $T_{L1_L}$                                & 0.33                                & 0.33                                         & 0.33  \\
      $T_{L1_{S}}$                              & 0.66                                & 0.66                                          & 0.66 \\
      $T_{L2}$                                 & $(\frac{64}{64}) + (\frac{64}{64}) = 2$ & $(2 \times \frac{64}{64}) + (\frac{64}{64}) = 3$  & $(4 \times \frac{64}{64}) + (\frac{64}{64}) = 5$  \\
      $T_{L3}$                                 & $(\frac{64}{64}) + (\frac{64}{64}) = 2$ & $(2 \times \frac{64}{64}) + (\frac{64}{64})= 3$  & $(4 \times \frac{64}{64}) + (\frac{64}{64}) = 5$ \\
      $T_{mem}$                                & $(\frac{64}{14}) + (\frac{64}{14}) = 9$ & $(\frac{64}{14}) + (\frac{64}{14}) = 9$           & ($\frac{64}{14}) + (\frac{64}{14}) = 9$  \\ \hline
      $T_{ECM}$                                & $14$                                & $16$                                                  &  $20$ \\ \hline
    \end{tabular}%
  \caption{ECM performance breakdown for packing $B_U$ for various strides for Intel Xeon Gold 6148 (Skylake-X).}
  \label{tab:ecm-perf-strides-skx}
\end{table}

When packing $A_{V^T}$, we make use of gather instructions, which means there is one {\tt VGATHERQPD}
and one store ({\tt VMOVAPD(store)}) being used per VL. It is hard to exactly model the behaviour
of the gather instruction since it seems to be fetching several cache lines together without incurring
the overhead for fetching each cache line individually. Our explanation is that as a result of page-based
fetching for the L3 cache, we can ignore the fact that multiple cache lines are being fetched and model that
as a single cache line fetch. With this assumption, we can state that $T_{L1_{L}} = 3$, $T_{L2_{S}} = 0.66$,
$T_{L2} = \frac{64}{64}$, $T_{L3} = 2 \times \frac{64}{64} = 2$ and $T_{mem} = \frac{64}{14} = 4.57$.

\subsubsection{$C_{MN}$ kernel performance analysis}
\label{sec:skx-cmn-kernels}

Each rank-1 update on the $C_{MN}$ kernel requires 1 {\tt VMOVAPD}, 8 {\tt VBROADCASTSD},
and 8 {\tt VFMADD231PD} instructions. Given that the data is always streamed from the
L1 cache, the only cost is $T_{L1_{L}} = T_{ECM} = 4.66$.

\subsection{Optimization on AMD EPYC 7502 with the ECM model}
\label{sec:optimize-epyc-ecm}

In order to keep the comparison between A64FX, AVX-512 and AVX2 fair, we use data sizes
that correspond to the vector length and its multiples for building the ECM models. Therefore,
in case of AVX2, the ranks used are 4, 8 and 16 which
correspond to the VL, twice the VL and four times the VL for each instruction. These
factors of the VL are proportional to the factors taken for A64FX and AVX512, for which
the ranks are 8, 16 and 32 as shown in Sec. \ref{sec:skx-optimize-ecm} and 
Sec. \ref{sec:optimize-fugaku-ecm} respectively. \rev{As shown in Sec.~\ref{sec:evaluate-blr-matvec},
the target application of the low rank multiplication is the matrix vector multiplication
routine for a block low rank matrix. The accuracy of the multiplication can be changed using
the admissibility condition, and we do not need to change the rank of the low rank 
matrices in order to modify the accuracy. Therefore, we test only for multiples
of the SIMD register length for each CPU.}

\subsubsection{Packing $A_{V^T}$ and $B_U$ performance analysis}
\label{sec:amd-packing-perf-analysis}

We can derive the ECM model by
using the throughput values from Table \ref{tab:cpu-instruction-table} and the machine model
from Table \ref{tab:machine-parameters-ecm}. $B_U$ packing takes one LOAD and one STORE
operation. Since the clock cycles depend on the stride, the theoretical performance for various
values of stride are shown in Table \ref{tab:ecm-perf-strides-amd}.

\begin{center}
\begin{table}[]
  \centering
    \begin{tabular}{|l|l|l|l|}
      \hline
      \multicolumn{1}{|c|}{\textbf{Variable}} & \multicolumn{1}{c|}{\textbf{Stride 4}} & \multicolumn{1}{c|}{\textbf{Stride 8}} & \multicolumn{1}{c|}{\textbf{Stride 16}} \\ \hline
      $T_{L1_L}$                                & 0.5                                          & 0.5                                                   &  0.5  \\
      $T_{L1_{S}}$                              & 0.75                                         & 0.75                                                   & 0.75 \\
      $T_{L2}$                                 & $\frac{32}{32} + 2 \times \frac{32}{32} = 3$ & $2 \times \frac{32}{32} + 2 \times \frac{32}{32} = 4$  & $4 \times \frac{32}{32} + 2 \times \frac{32}{32} = 6$ \\
      $T_{L3}$                                 & $\frac{32}{32} + 2 \times \frac{32}{32} = 3$ & $2 \times \frac{32}{32} + 2 \times \frac{32}{32} = 4$  & $4 \times \frac{32}{32} + 2 \times \frac{32}{32} = 6$ \\
      $T_{mem}$                                & $\frac{32}{16} + \frac{32}{16} = 4$          & $\frac{32}{16} + \frac{32}{16} = 4$                     & $\frac{32}{16} + \frac{32}{16} = 4$  \\ \hline
      $T_{ECM}$                                & $4$                                          & $4$                                                     &  $6$ \\ \hline
    \end{tabular}%
  \caption{ECM performance breakdown for packing $B_U$ for various strides for a
    single thread on the AMD EPYC 7502. All measurements are reported in number of clock cycles.}
  \label{tab:ecm-perf-strides-amd}
\end{table}
\end{center}

\subsubsection{$C_{MN}$ kernel performance analysis}
\label{sec:amd-cmn-perf-analysis}

The $C_{MN}$ kernel in this case is very similar to that in Intel. The difference being
that the VL is limited to 4 due to the AVX2 instruction set. Therefore, the rank-1 update
in this case is for a 4x4 matrix block, unlike the 8x8 matrix block for the Intel.
For each rank-1 update, we use 4 {\tt VBROADCASTSD}, 4 {\tt FMADD231PD} and 1 {\tt VMOVAPD}
instructions. This amounts to $T_{L1_{L}} = 3.5$. Since the data is directly streamed from
the L1 cache, all other terms in the ECM equation are $0$ and therefore $T_{ECM} = 3.5$. 

%% file: experiments.tex
\section{Experimental Evaluation}
\label{sec:exp-evaluation}

We perform experiments using the nodes and compiler flags listed in Table \ref{tab:node-compile-config}. 
The hardware specification 
of the CPU within each node can be found in Table \ref{tab:machine-parameters-ecm}.
Our method works for any kind of data. However, for these experiments
we use randomly generated entries following a normal distribution in order to accurately evaluate all of our 
test matrices.
Since we are only working with low rank multiplication, the data 
within the low 
rank blocks does not
affect our results. All tests are performed
using double precision floating point numbers.

\begin{table}[tpb]
    \begin{tabular}{|c|c|c|c|}
    \hline
    \multicolumn{1}{|l|}{}                                                  & \textbf{AMD}                                                                                                & \textbf{Fujitsu}                                                                                                             & \textbf{Intel}                                                                                                                \\ \hline
    CPUs                                                                    & 2 x AMD EPYC 7502                                                                                           & 1 x A64FX                                                                                                                    & 2 x Intel Skylake-X 6148                                                                                                      \\ \hline
    Memory                                                                  & 2 x 256 GiB                                                                                                 & 1 x 32 GiB (HBM)                                                                                                             & $2 \times 192$ GiB                                                                                                            \\ \hline
    \begin{tabular}[c]{@{}c@{}}NUMA \\ configuration\end{tabular}           & 4/CPU                                                                                                       & 4/CPU                                                                                                                        & 1 NUMA node/CPU                                                                                                               \\ \hline
    Compiler                                                                & g++ 7.5.0                                                                                                   & FCC 4.5.0 tcsds-1.2.31                                                                                                       & g++ 7.4.0                                                                                                                     \\ \hline
    \begin{tabular}[c]{@{}c@{}}Compile \\ options\end{tabular}              & \begin{tabular}[c]{@{}c@{}}-Wall -fopenmp -O3 \\ -Ofast -mavx2\\ -funroll-loops \\ -masm=intel\end{tabular} & \begin{tabular}[c]{@{}c@{}}-O3 -Nfjomplib \\ -fopenmp \\ -Kfast,zfill\\ -Kopenmp \\ -Ksimd\_reg\_size=512 \end{tabular}      & \begin{tabular}[c]{@{}c@{}}-Wall -fopenmp -O3 \\ -Ofast \\ -march=skylake-avx512\\ -masm=intel\end{tabular}                   \\ \hline
    Math library                                                            & AMD BLIS 3.0.0                                                                                              & Fujitsu SSL-2                                                                                                                & Intel MKL 2020.4                                                                                                              \\ \hline
    \begin{tabular}[c]{@{}c@{}}Math library \\ linking options\end{tabular} & \begin{tabular}[c]{@{}c@{}}libblis-mt.a -lgomp \\ -lpthread -lm -ldl\end{tabular}                           & \begin{tabular}[c]{@{}c@{}}-Kopenmp -Nfjomplib \\ -lfjlapacksve\end{tabular}                                                 & \begin{tabular}[c]{@{}c@{}}-lmkl\_intel\_ilp64 \\ -lmkl\_gnu\_thread \\ -lmkl\_core -lgomp \\ -lpthread -lm -ldl\end{tabular} \\ \hline
    \begin{tabular}[c]{@{}c@{}}TRIAD \\ peak (Gb/s) \end{tabular}           &  195                                                                                                        & 840                                                                                                                          & 150                                                                                                                           \\ \hline
    \begin{tabular}[c]{@{}c@{}}DGEMM \\ peak (GFLOPS) \end{tabular}         & 2184                                                                                                        & 2828                                                                                                                         & 2621                                                                                                                          \\ \hline
    \end{tabular}
    \caption{Machine architecture and the corresponding configuration used in our experiments. 
    The Intel node is a single node of the ABCI supercomputer and the
    Fujitsu node is a single node of the FUGAKU supercomputer. The AMD 
    node is a stand-alone SMP machine.}                                                               
    \label{tab:node-compile-config}
\end{table}

\begin{equation}
    GFLOPS = \frac{batch\_size \times (4 \times rank^3 + 2 \times rank^2 \times block\_size)}{time(s)} \times 10^{-9}
\label{eq:exp-gflops-calculation}
\end{equation}

The GFLOPS, where shown, are calculated as presented in Eq. \ref{eq:exp-gflops-calculation}. Bandwidth calculation
depends on the cache overlapping displayed by the particular CPU and will be specified 
where necessary. We compare the bandwidth of each problem size with the STREAM TRIAD 
bandwidth for a given number of physical cores in order to demonstrate the `ideal' 
usable bandwidth vs. what is actually realized.

Experiments are performed using the full node available. For all cases except the A64FX
we use the OpenMP configuration as \texttt{OMP\_PLACES=cores} and \texttt{OMP\_PROC\_BIND=close}.
We run all tests using \texttt{numactl} using the
\texttt{--membind=all} configuration and
set \texttt{--physcpubind} to bind distinct physical cores.
The Fujitsu runtime in FUGAKU imposes a slightly different binding process than other machines.
Spawning a single process per CMG (i.e. one process per 12 cores) is the recommended configuration for using all NUMA nodes with strictly local access. Thus, we run 4 MPI processes per node for the A64FX tests.

\subsection{Evaluation on the Fujitsu node}
\label{sec:evaluation-fujitsu-a64fx}

\begin{figure}
  \centering
  \subfigure[Block size 512]{
    \includegraphics[width=0.31\linewidth]{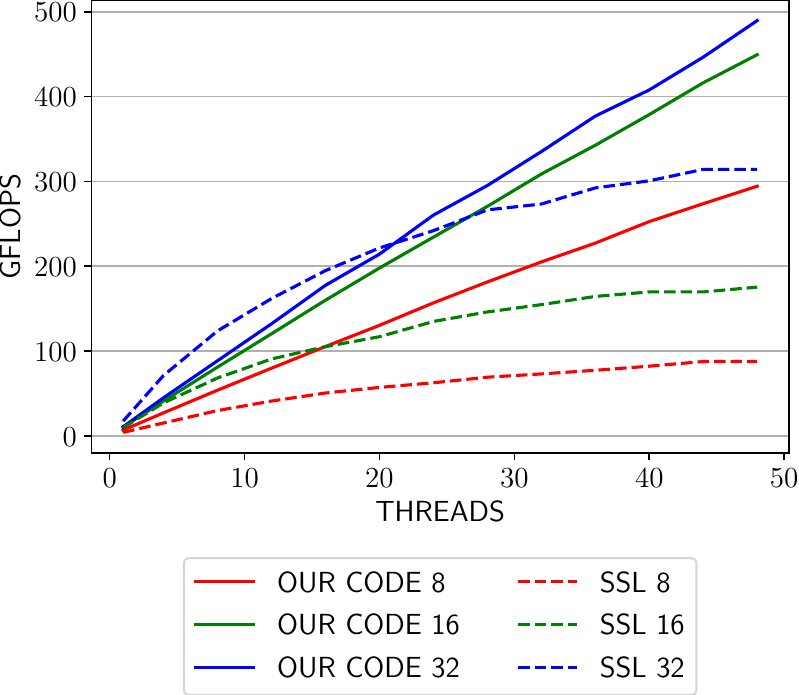}
    \label{fig:a64fx-gflops-batch-20k-512}
  }
  \subfigure[Block size 1024]{
    \includegraphics[width=0.31\linewidth]{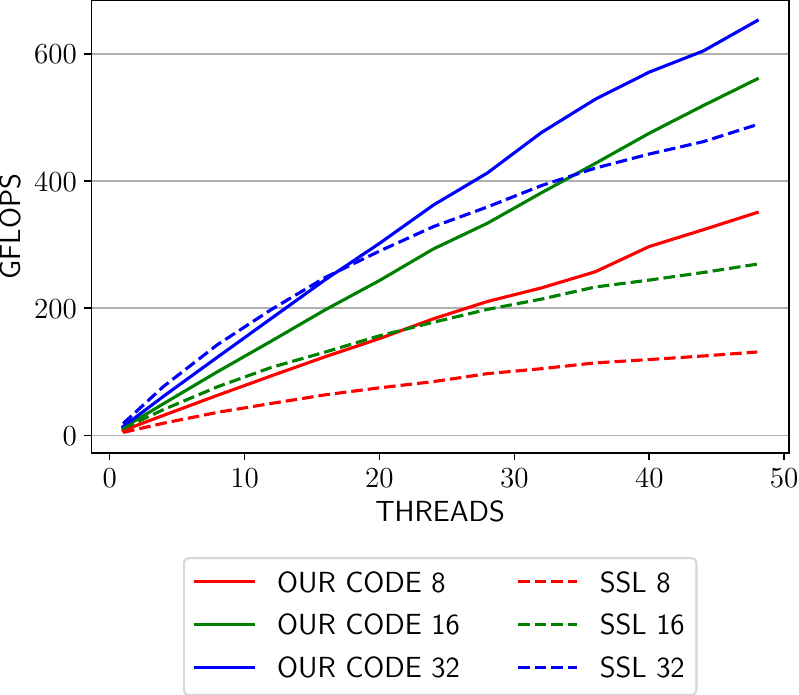}
    \label{fig:a64fx-gflops-batch-20k-1024}
  }
  \subfigure[Block size 2048]{
    \includegraphics[width=0.31\linewidth]{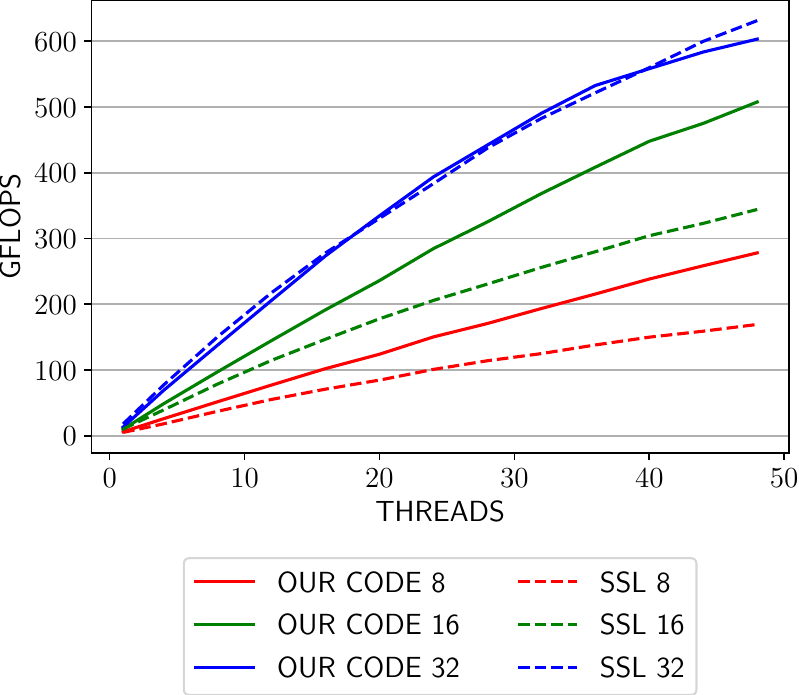}
    \label{fig:a64fx-gflops-batch-20k-2048}
  }
  \caption{Comparison of performance in GFLOPS vs. non-batched SSL-2 routines for 
  Fujitsu A64FX. The batch size is kept constant at 20,000 for all the tests. The
  legend indicates the rank for each plot.}
  \label{fig:a64fx-gflops-batch-20k}
\end{figure}

\begin{figure}
  \centering
  \subfigure[Block size 512]{
    \includegraphics[width=0.31\linewidth]{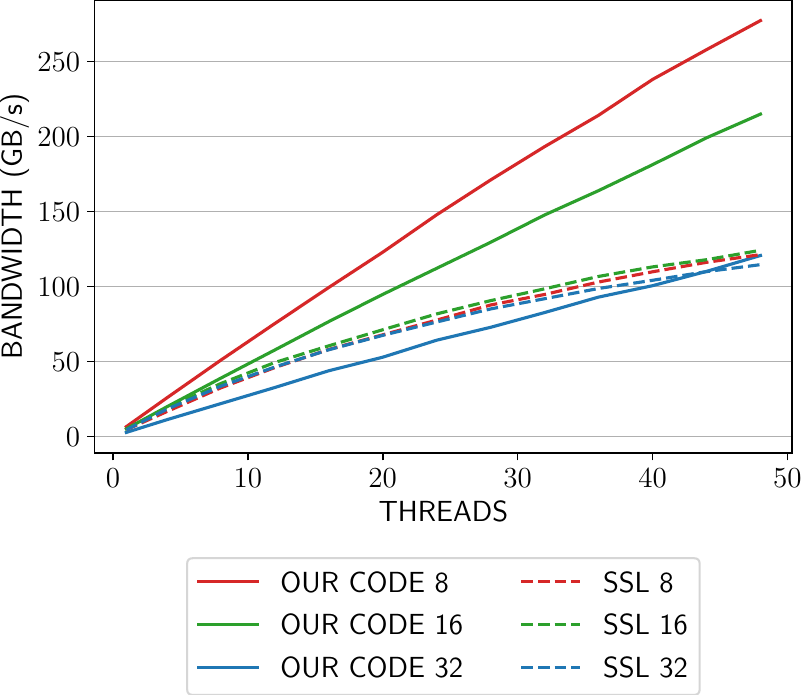}
    \label{fig:a64fx-bandwidth-util-block-512}
  }
  \subfigure[Block size 1024]{
    \includegraphics[width=0.31\linewidth]{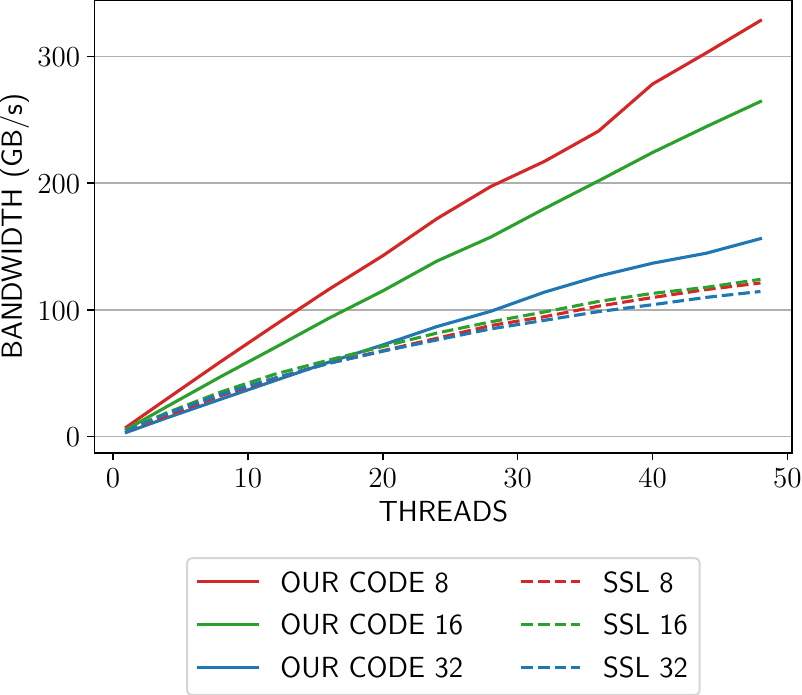}
    \label{fig:a64fx-bandwidth-util-block-1024}
  }
  \subfigure[Block size 2048]{
    \includegraphics[width=0.31\linewidth]{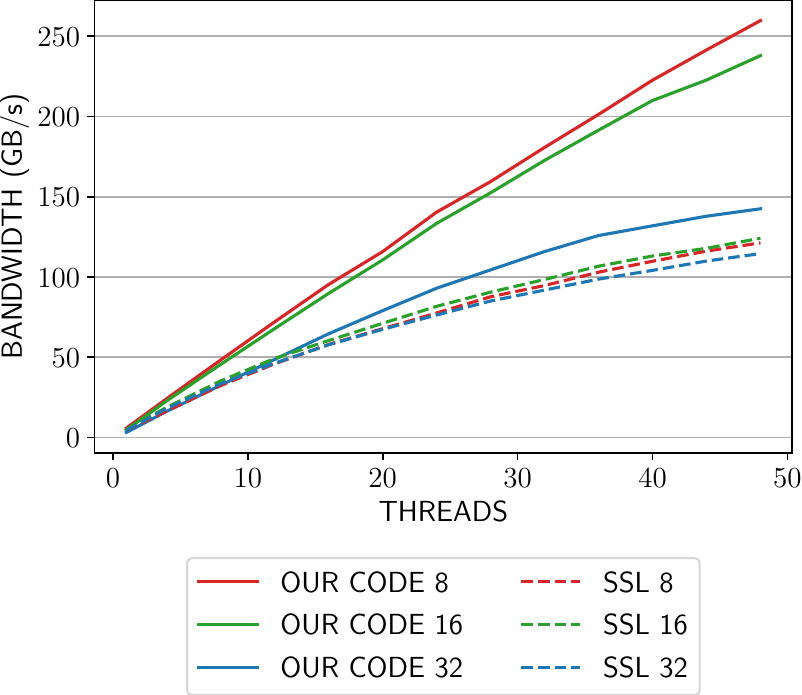}
    \label{fig:a64fx-bandwidth-util-block-2048}
  }
  \caption{Bandwidth utilization for varying ranks and block sizes on Fujitsu A64FX. The legend shows the rank for each plot. The bandwidth is calculated using Eq. \ref{eq:amd-exp-bw-calculation}. The peak STREAM TRIAD
  bandwidth for the A64FX is about 840 GB/s.}
  \label{fig:a64fx-bandwidth-util}
\end{figure}

The utilization on the Fujitsu A64FX is shown in Fig. \ref{fig:a64fx-gflops-batch-20k}. It
can be seen that our code outperforms Fujitsu's SSL-2 library by a wide margin except for
one case using rank $32$ and block size $2048$ when not using all the cores in the CPU.
The bandwidth utilization plots in Fig. \ref{fig:a64fx-bandwidth-util} show that, although
the GFLOPS utilization for rank $32$ is consistently higher than other ranks, the bandwidth
utilization is lower. This result indicates that the rank $32$ case is in fact compute bound and
not memory bound when using smaller ranks.

\begin{equation}
    BW (GiB/s) = \frac{batch\_size \times (2 \times rank^2 + 2 \times rank \times block\_size) \times sizeof(double) \times 2^{-30}}{time(s)}
    \label{eq:amd-exp-bw-calculation}
\end{equation}

Lack of linear strong scaling can be seen when using rank $8$ and block size $512$ in 
Fig. \ref{fig:a64fx-gflops-batch-20k}, which gradually improves as the block size is
increased. This effect is not observed for other ranks that show almost uniform linear
scaling. The reason for this can be seen from the bandwidth utilization plot in
Fig. \ref{fig:a64fx-bandwidth-util}, where the usage of the bandwidth is proportional
to the GFLOPS utilization. The bandwidth is calculated as shown in Eq.
\ref{eq:amd-exp-bw-calculation}. The bandwidth utilization is lesser than 
for rank $16$ for
the same block sizes since packing smaller skinny matrices individually into the L1
cache does not lead to optimal bandwidth utilization. Increasing the number of skinny
matrices that are packed into the L1 cache during a single iteration of Loop 1 in
Algorithm \ref{alg:low-rank-mult-six-nest} might be a way to solve this problem.

\begin{figure}
  \centering
  \subfigure[Block size 512]{
    \includegraphics[width=0.31\linewidth]{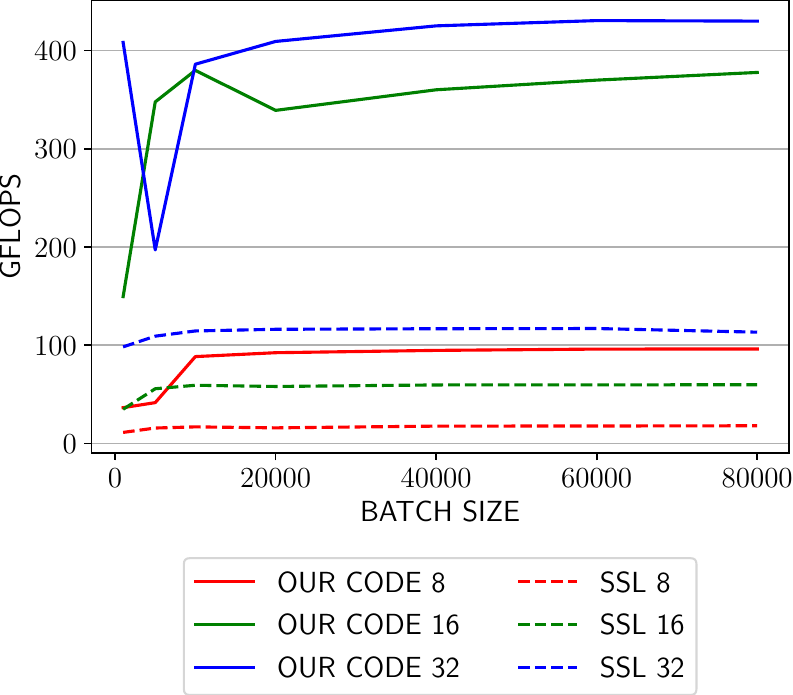}
    \label{fig:a64fx-threads-const-512}
  }
  \subfigure[Block size 1024]{
    \includegraphics[width=0.31\linewidth]{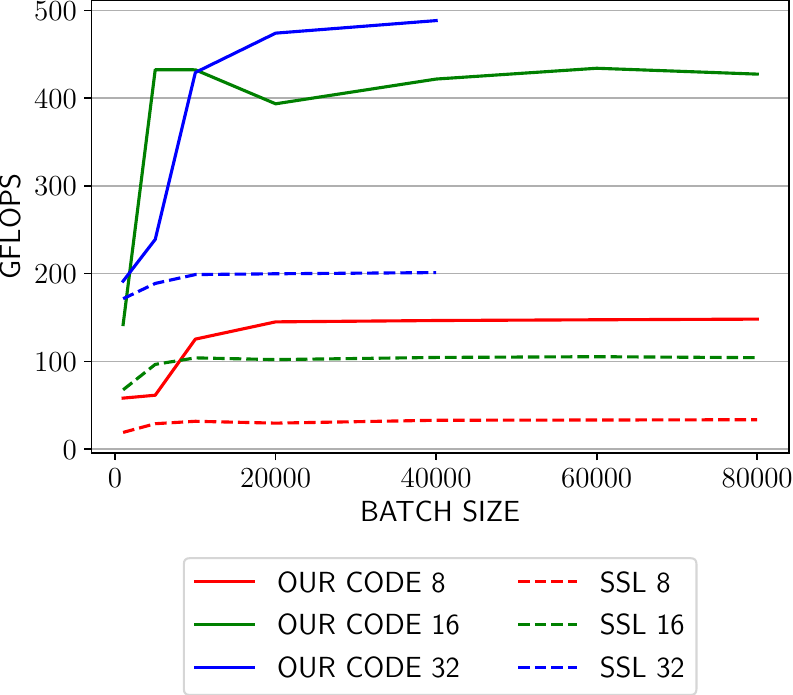}
    \label{fig:a64fx-threads-const-1024}
  }
  \subfigure[Block size 2048]{
    \includegraphics[width=0.31\linewidth]{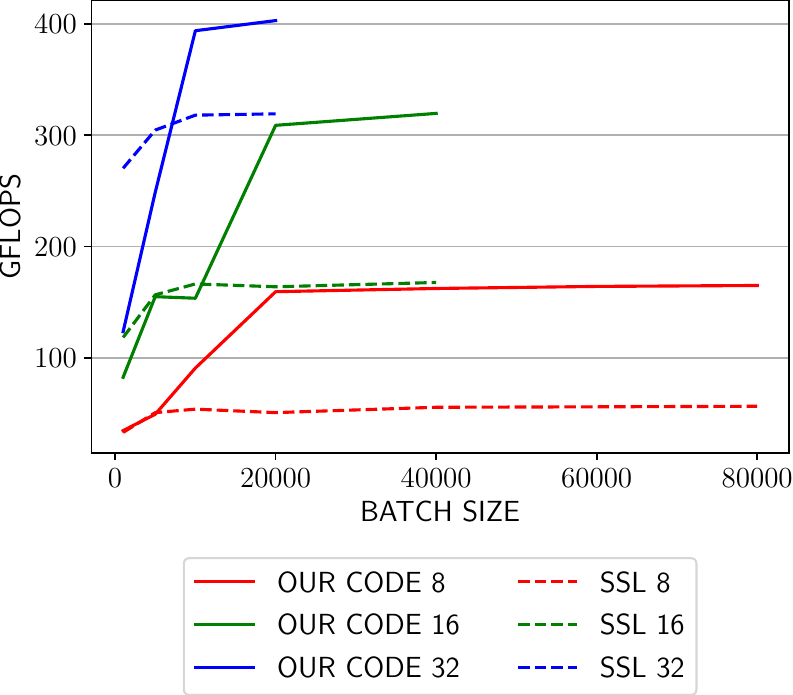}
    \label{fig:a64fx-threads-const-2048}
  }
  \caption{Performance for various block sizes and ranks when the number of threads is
    constant at 48 for varying batch sizes for Fujitsu A64FX.}
  \label{fig:a64fx-threads-const-batch-variation}
\end{figure}

When using rank $32$ and block size $2048$ in Fig. \ref{fig:a64fx-bandwidth-util-block-2048},
it can be seen that the bandwidth utilization of SSL-2 and our code is almost the same unless
all 4 NUMA nodes within the CPU are active, which happens after 36 threads are active.
Fig. \ref{fig:a64fx-threads-const-batch-variation} shows the variation of the performance in
GFLOPS as the number of threads is kept constant at 48 and batch size is varied. Our code consistently
outperforms SSL-2, and there is minimal variation in the GFLOPS as the batch size is changed. Some results for the Fujitsu node could not be
reported due to the limited 32 GiB HBM. However, it can be seen that each plot plateaus
before we run out of memory, and therefore we can assume that performance will not degrade
if there were more memory.

\begin{figure}[tpb]
    \centering
    \includegraphics[width=\linewidth]{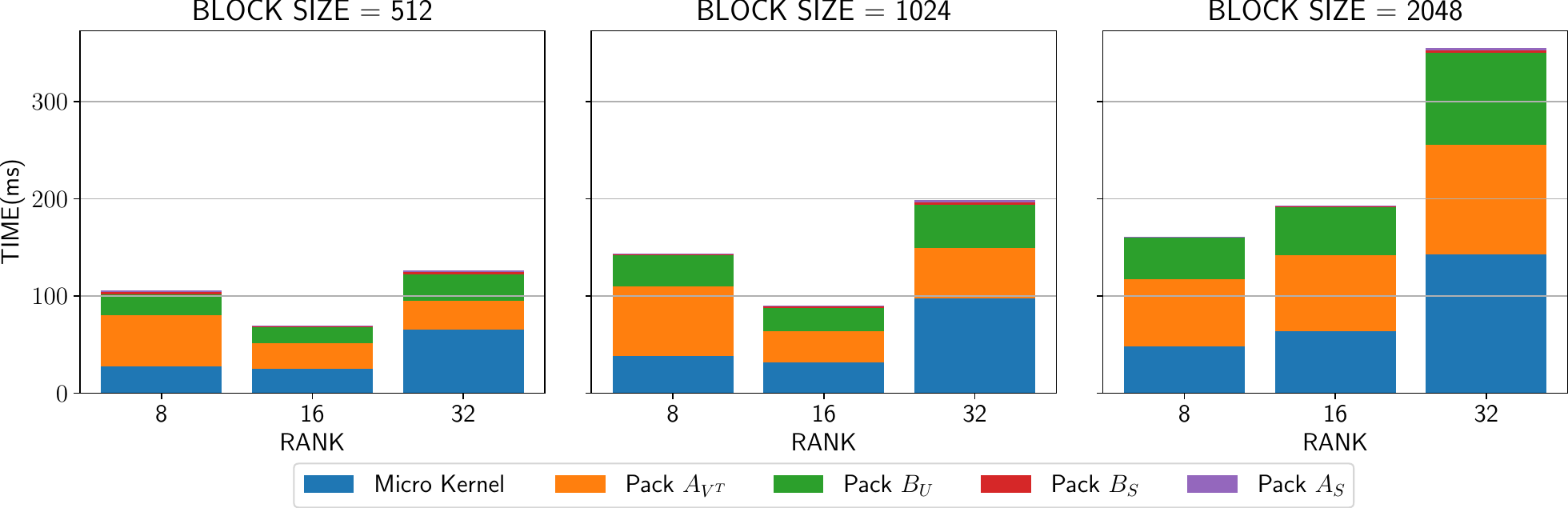}
    \caption{Performance breakdown for various parts of the computation for the 
        Fujitsu node using 48 threads and a constant batch size of 20000.}
    \label{fig:perf-breakdown-a64fx}
\end{figure}

Fig. \ref{fig:perf-breakdown-a64fx} shows the performance breakdown for
each kernel of the computation on the A64FX when using 48 physical cores.

\rev{
Table~\ref{tab:a64fx-higher-rank} shows that as the rank increases beyond 96, we 
can observe SSL showing better performance than our code. This can be attributed 
to the fact that the algorithm becomes more compute bound. The performance actually
drops below that of rank 32, which can be attributed to the fact that better bandwidth
utilization as shown in Fig.~\ref{fig:a64fx-bandwidth-util} becomes harder as a lesser
number of small matrices can be packed into the L1 cache. This is one of the drawbacks
of relying on packing intermediate products into the SIMD registers.

\begin{table}[]
\begin{tabular}{|l|l|l|l|l|}
\hline
\multicolumn{1}{|c|}{\textbf{Batch size}} & \multicolumn{1}{c|}{\textbf{Block size}} & \multicolumn{1}{c|}{\textbf{Rank}} & \multicolumn{1}{c|}{\textbf{SSL}} & \multicolumn{1}{c|}{\textbf{OUR CODE}} \\ \hline
20000  & 512 & 96 & 370.62 & 207.55 \\ \hline
20000   & 1024 & 96 & 598.93 & 343.912    \\ \hline
20000  & 2048 & 96 & 851.74 & 521.93 \\ \hline
\end{tabular}
\caption{\rev{Performance of our code vs. SSL for larger rank on the Fujitsu node using 48 
physical cores reported in GFLOPS. SSL shows better performance as the computation becomes 
more compute bound.}}
\label{tab:a64fx-higher-rank}
\end{table}
}

\subsection{Evaluation on the Intel node}
\label{sec:evaluation-skylake-x}

\begin{figure}
  \centering
  \subfigure[Block size 512]{
    \includegraphics[width=0.31\linewidth]{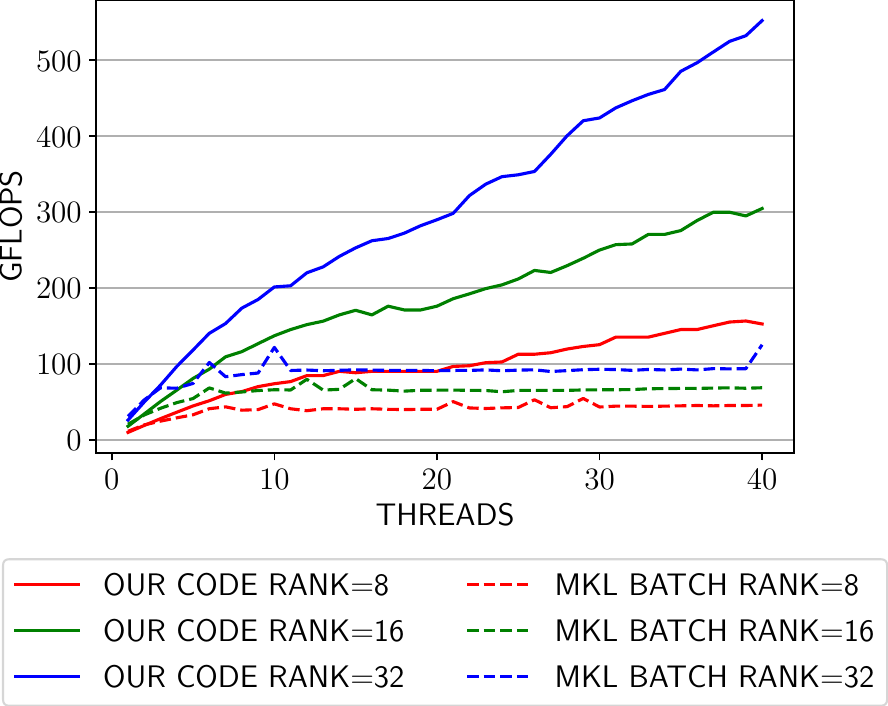}
    \label{fig:skx-gflops-batch-20k-512}
  }
  \subfigure[Block size 1024]{
    \includegraphics[width=0.31\linewidth]{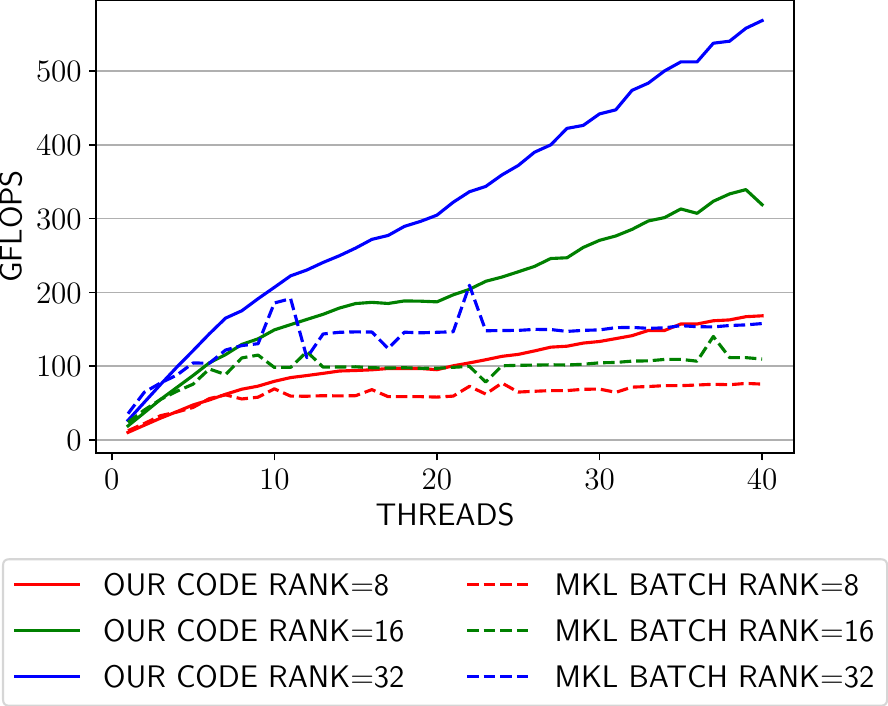}
    \label{fig:skx-gflops-batch-20k-1024}
  }
  \subfigure[Block size 2048]{
    \includegraphics[width=0.31\linewidth]{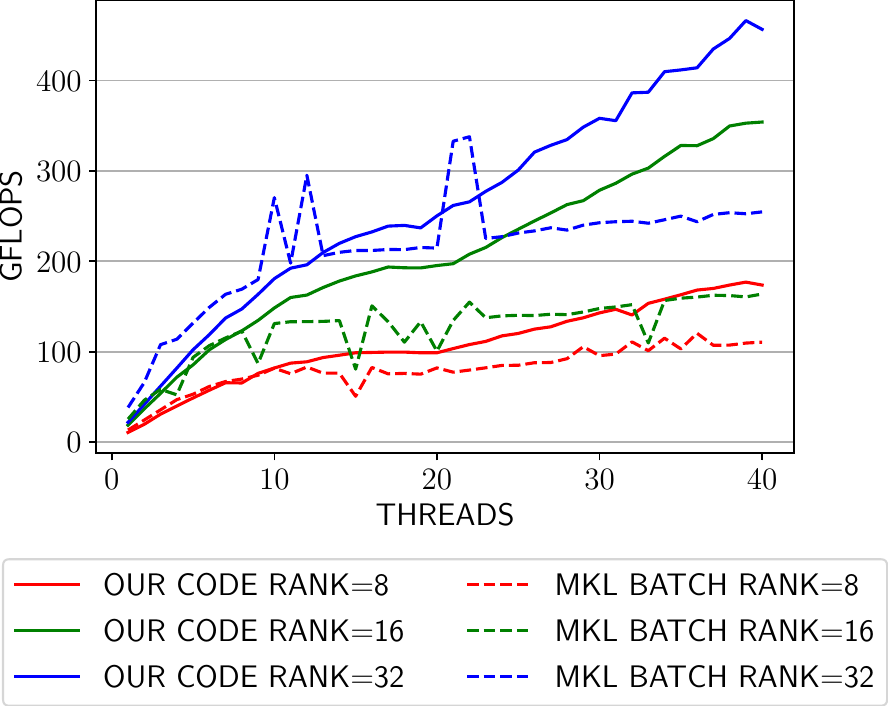}
    \label{fig:skx-gflops-batch-20k-2048}
  }
  \caption{Comparison of performance in GFLOPS vs. batched MKL 
    routines for Intel Skylake-X. The batch size
    is kept constant at 20,000 for all the tests.}
  \label{fig:skx-gflops-batch-20k}
\end{figure}

Fig. \ref{fig:skx-gflops-batch-20k} shows the utilization in GFLOPS when the batch size is kept
constant at 20,000 for a varying number of threads, block sizes, and ranks. It can be seen that
our approach shows almost perfect scaling with respect to the number of threads whereas batched
MKL routines stop scaling after approximately 10 physical cores have been utilized for all problem
cases. We do not report findings for non-batched MKL routines since they show the least competitive
performance for all problem cases. 

\begin{figure}
  \centering
  \subfigure[Block size 512]{
    \includegraphics[width=0.31\linewidth]{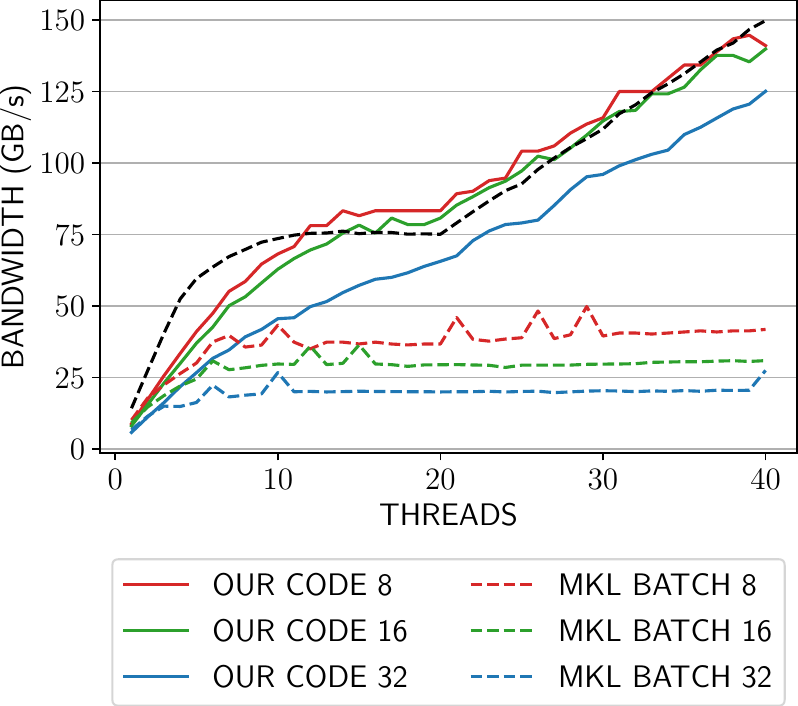}
    \label{fig:skx-bandwidth-util-block-512}
  }
  \subfigure[Block size 1024]{
    \includegraphics[width=0.31\linewidth]{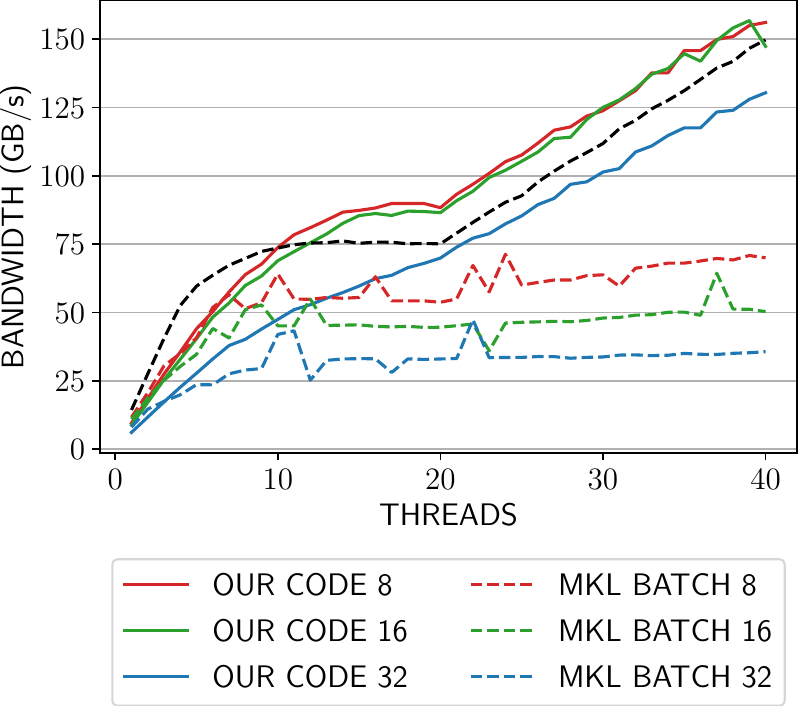}
    \label{fig:skx-bandwidth-util-block-512}
  }
  \subfigure[Block size 2048]{
    \includegraphics[width=0.31\linewidth]{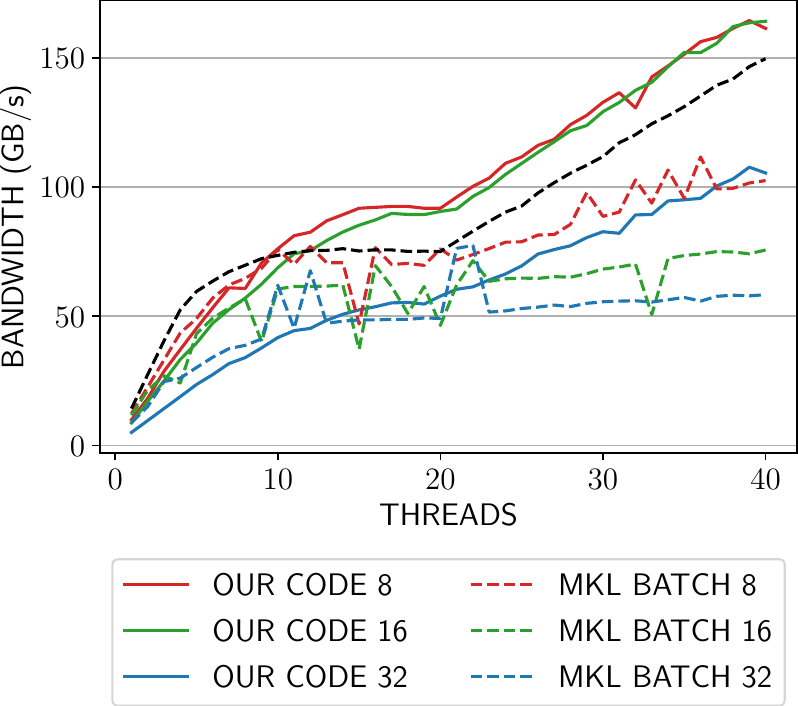}
    \label{fig:skx-bandwidth-util-block-512}
  }
  \caption{Comparison of bandwidth utilization for varying ranks,
  threads and block sizes on Intel Xeon Gold 6148 (Skylake-X) for
  a constant batch size of 20,000. The black dashed line denotes
    the STREAM TRIAD bandwidth for the given number of threads.}
  \label{fig:skx-bandwidth-util}
\end{figure}

The improved strong scaling can be attributed to the fact that our approach is able to saturate the maximum available bandwidth much better than batched MKL, as can be seen in Fig. \ref{fig:skx-bandwidth-util}, which shows the bandwidth utilization in GiB/second with a constant batch size of 20,000 and varying number of threads, block sizes and ranks. This is as a result of our unique packing strategy.

\begin{figure}
  \centering
  \subfigure[Block size 512]{
    \includegraphics[width=0.31\linewidth]{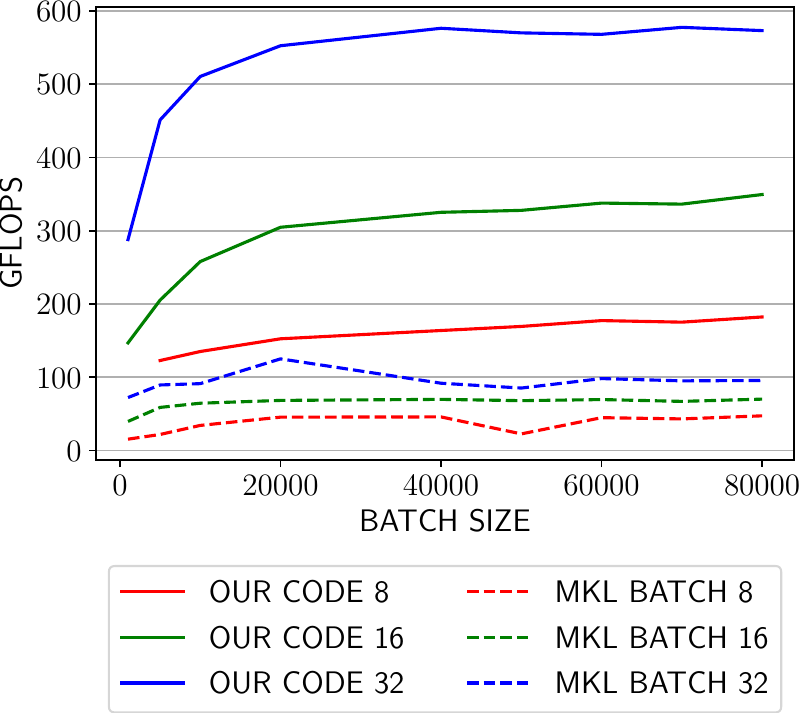}
    \label{fig:skx-threads-const-512}
  }
  \subfigure[Block size 1024]{
    \includegraphics[width=0.31\linewidth]{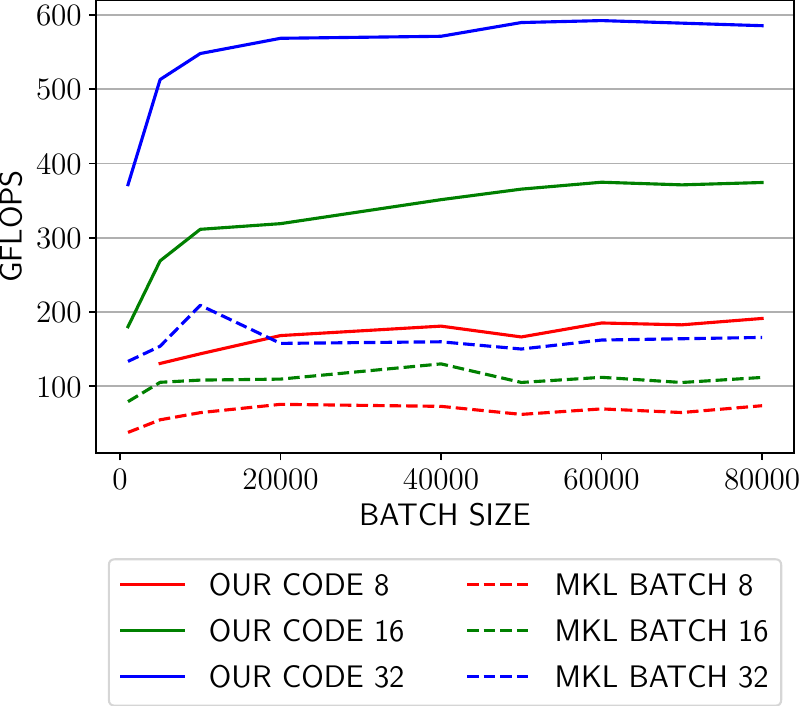}
    \label{fig:skx-threads-const-1024}
  }
  \subfigure[Block size 2048]{
    \includegraphics[width=0.31\linewidth]{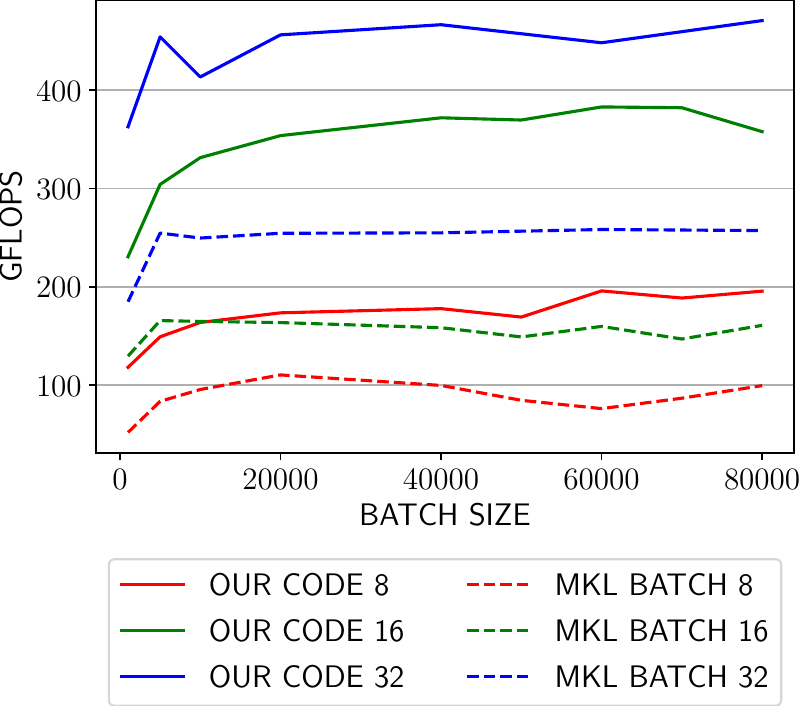}
    \label{fig:skx-threads-const-2048}
  }
  \caption{Performance for various block sizes and ranks when the number of threads is
    constant at 40 for the Intel Xeon Gold 6148 (Skylake-X).}
  \label{fig:skx-threads-const}
\end{figure}

The bandwidth numbers shown in Fig. \ref{fig:skx-bandwidth-util} are calculated as shown
in Eq. \ref{eq:skx-exp-bw-calculation}. We use the term $3 \times rank^2$ in order to
account for the write of the result matrix and reads of two small matrices. As the ECM
model for Intel Skylake-X proves in Sec. \ref{sec:ecm-overlap-hypothesis}, the reads
and writes for this CPU are completely non-overlapping and the write term must be added
into the bandwidth calculation.

\begin{equation}
    BW (GiB/s) = \frac{batch\_size \times (3 \times rank^2 + 2 \times rank \times block\_size) \times sizeof(double) \times 2^{-30}}{time(s)}
    \label{eq:skx-exp-bw-calculation}
\end{equation}

Fig. \ref{fig:skx-threads-const} shows the performance when changing the batch size,
block size and rank and keeping the number of threads constant at 40. It can be 
seen that both our implementation, and batched MKL show almost constant performance
irrespective of the batch size. This, combined with Fig. \ref{fig:skx-gflops-batch-20k}
shows that the scaling of the method is primarily limited 
by the available bandwidth, rank and block size. Additionally, increasing the 
batch size does not have any effect on the performance.

\begin{figure}
    \centering
    \includegraphics[width=\linewidth]{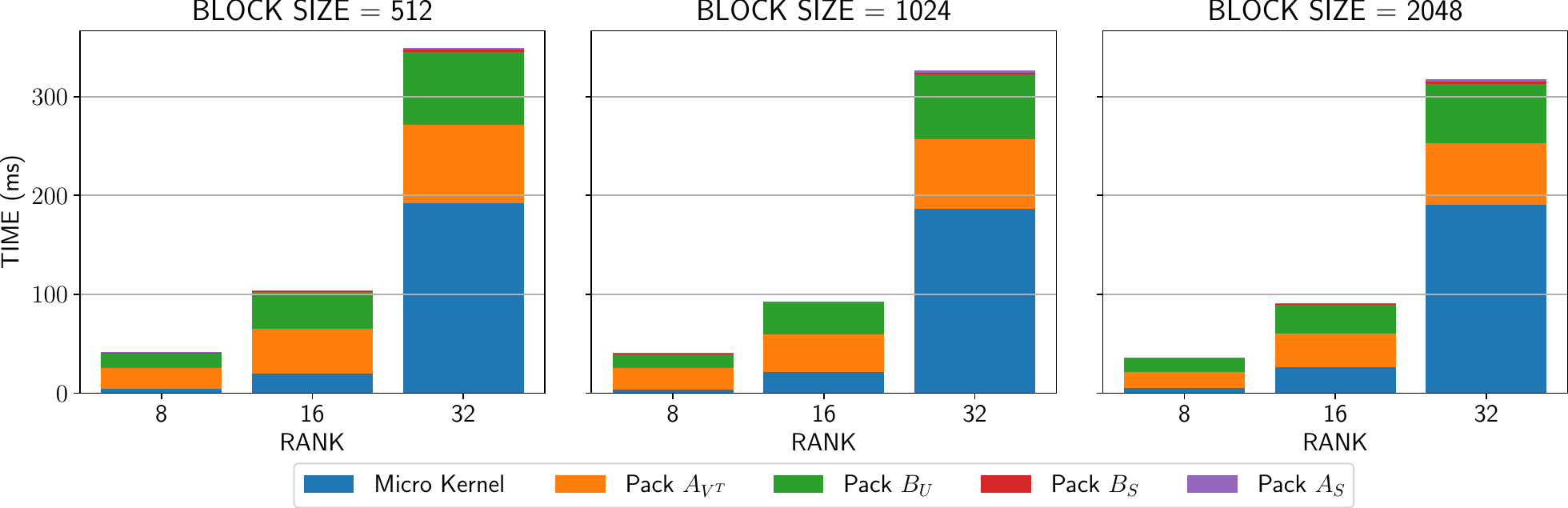}
    \caption{Performance breakdown for various parts of the computation for the Intel node
        using 40 threads and a batch size of 20000.}
    \label{fig:perf-breakdown-intel}
\end{figure}

Fig.~\ref{fig:perf-breakdown-intel} shows the time spent in the micro kernel (i.e. performing actual computation) vs. the time spent in packing data into the caches. This graph is a snapshot of the experiment with a batch size of 20,000 from Fig.~\ref{fig:skx-threads-const}. 

As the rank increases beyond 128, we can observe MKL batched showing better performance than our code. Table~\ref{tab:intel-rank-128} shows the difference in performance as the computation becomes more compute bound than memory bound.

\begin{table}[]
\begin{tabular}{|l|l|l|l|l|}
\hline
\multicolumn{1}{|c|}{\textbf{Batch size}} & \multicolumn{1}{c|}{\textbf{Block size}} & \multicolumn{1}{c|}{\textbf{Rank}} & \multicolumn{1}{c|}{\textbf{MKL BATCH}} & \multicolumn{1}{c|}{\textbf{OUR CODE}} \\ \hline
20000   & 512   & 128 & 262.281  & 339.849                                \\ \hline
20000  & 1024  & 128 & 400.602   & 361.89  \\ \hline
20000                                     & 2048                                     & 128                                & 554.721                                 & 319.161                                \\ \hline
\end{tabular}
\caption{Performance of our code vs. MKL batched for larger rank on the Intel node 
using 40 physical cores in GFLOPS. Batched MKL shows better performance as the 
computation becomes more compute bound.}
\label{tab:intel-rank-128}
\end{table}

\subsection{Evaluation on the AMD node}
\label{sec:evaluation-amd-epyc}

\begin{figure}[htbp]
  \centering
  \subfigure[Block size 512]{
    \includegraphics[width=0.31\linewidth]{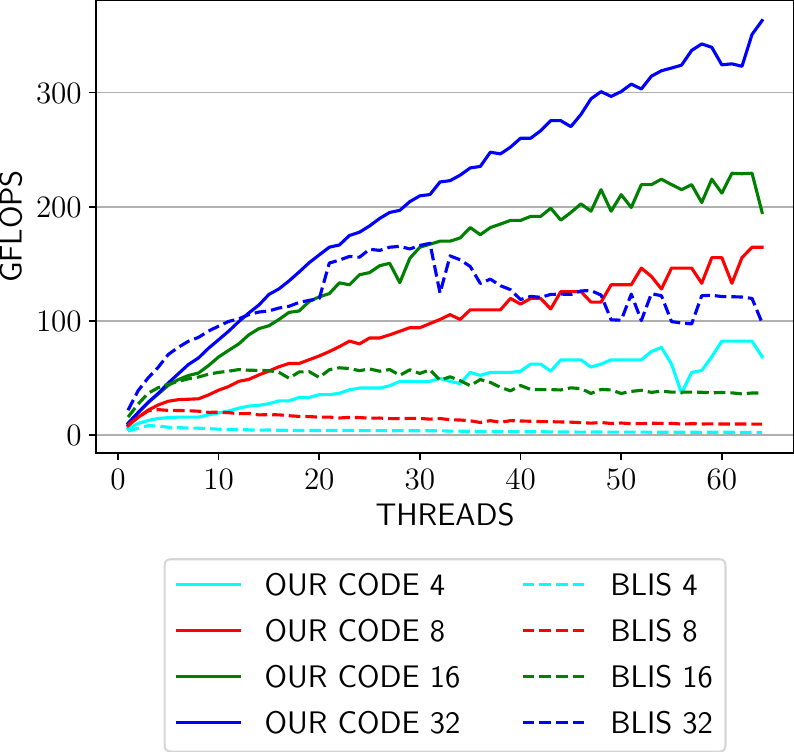}
    \label{fig:amd-gflops-batch-20k-512}
  }
  \subfigure[Block size 1024]{
    \includegraphics[width=0.31\linewidth]{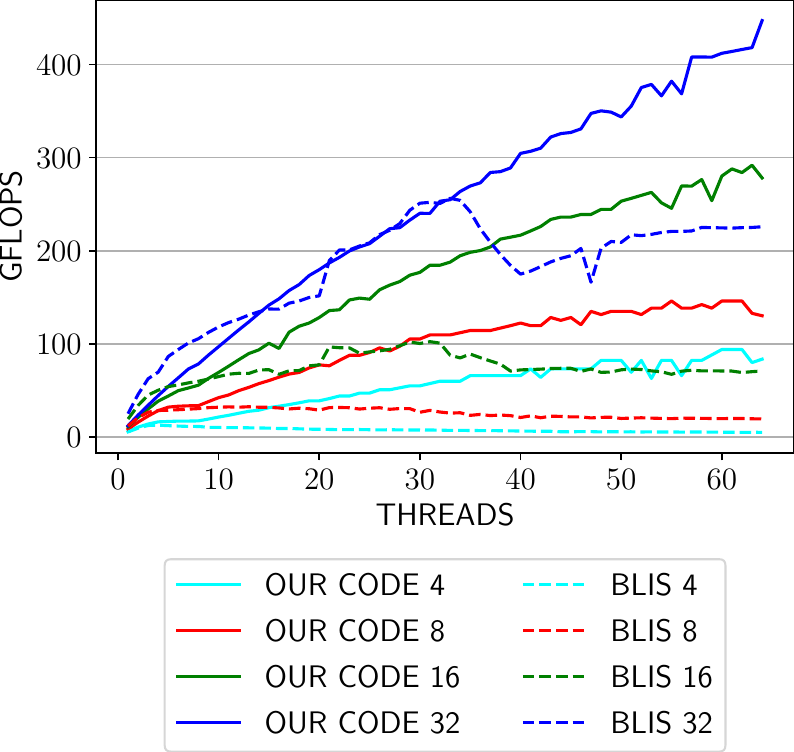}
    \label{fig:amd-gflops-batch-20k-1024}
  }
  \subfigure[Block size 2048]{
    \includegraphics[width=0.31\linewidth]{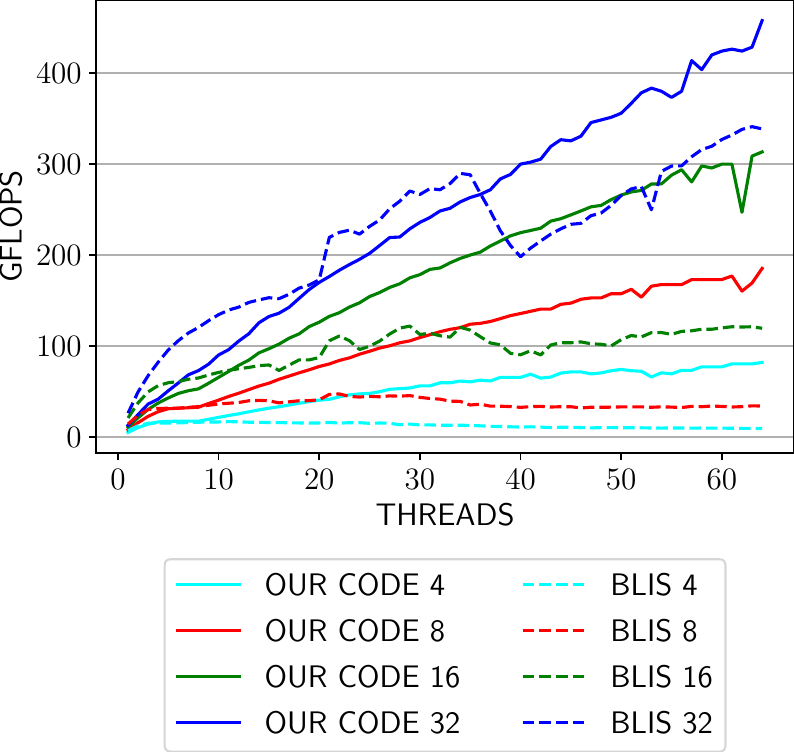}
    \label{fig:amd-gflops-batch-20k-2048}
  }
  \caption{Comparison of performance in GFLOPS vs. non-batched AMD
  BLIS routines for 
  AMD EPYC 7502. The batch size is kept constant at 20,000 for 
  all the tests. The
  legend indicates the rank for each plot. }
  \label{fig:amd-gflops-batch-20k}
\end{figure}

As shown in Table \ref{tab:machine-parameters-ecm}, the vector length of the AMD CPU is
4, whereas that of the Intel and Fujitsu CPUs is 8. Therefore, we show benchmarks for
rank $4$ in case of the AMD chip as well as for rank 8, 16 and 32 as shown for the others.
Fig. \ref{fig:amd-gflops-batch-20k} shows the utilization as calculated using 
Eq. \ref{eq:exp-gflops-calculation}, which shows that our implementation is able
to outperform the vendor optimized AMD BLIS 3.0.0 BLAS implementation by a wide margin
when a sufficient number of physical cores are active. The performance improves as we increase the rank, as expected, since memory bandwidth becomes less of a bottleneck
as more data becomes available to keep the FMA units of the CPU busy.

\begin{figure}[htbp]
  \centering
  \subfigure[Block size 512]{
    \includegraphics[width=0.31\linewidth]{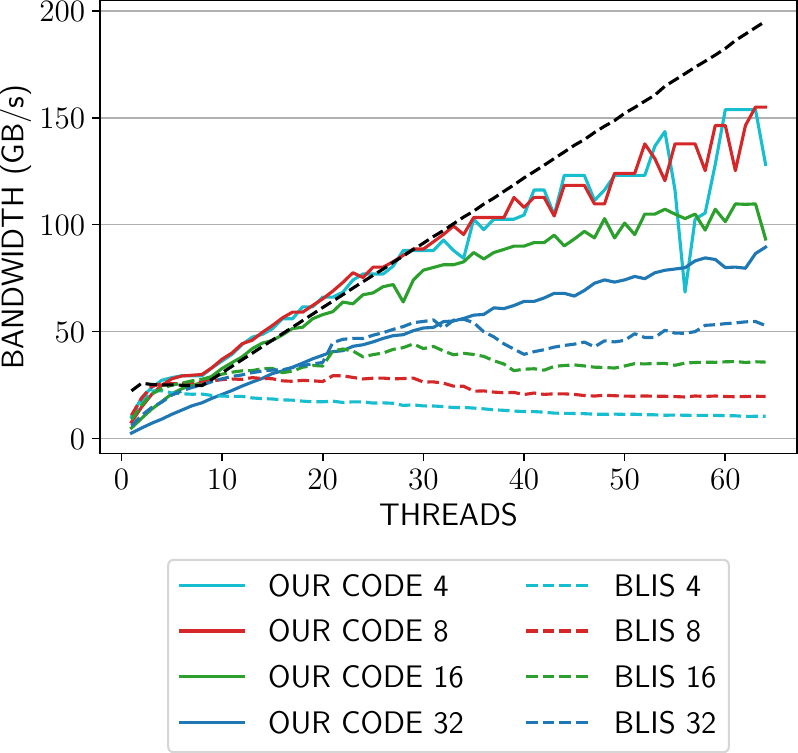}
    \label{fig:amd-bandwidth-util-block-512}
  }
  \subfigure[Block size 1024]{
    \includegraphics[width=0.31\linewidth]{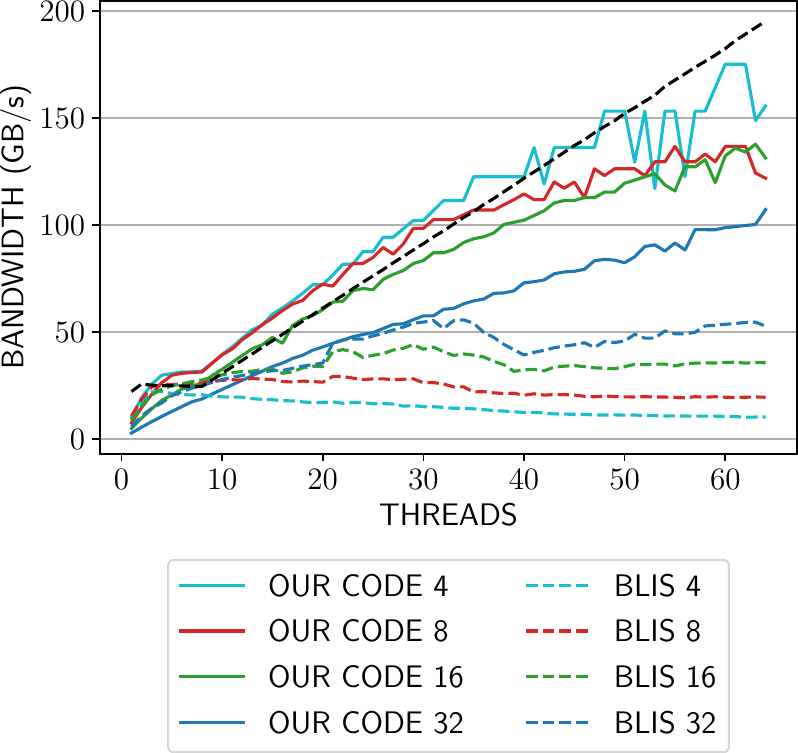}
    \label{fig:amd-bandwidth-util-block-512}
  }
  \subfigure[Block size 2048]{
    \includegraphics[width=0.31\linewidth]{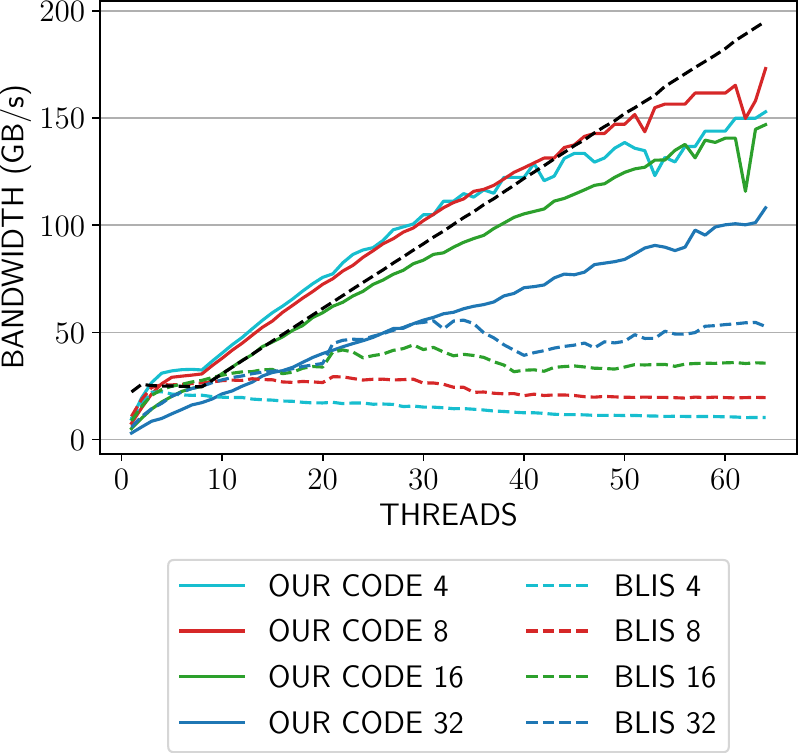}
    \label{fig:amd-bandwidth-util-block-512}
  }
  \caption{Bandwidth utilization for varying ranks and block sizes on AMD EPYC 7502. The legend shows the rank for each plot. The black 
  dashed line shows the STREAM TRIAD bandwidth for the 
  given number of threads.}
  \label{fig:amd-bandwidth-util}
\end{figure}

\begin{figure}[htbp]
  \centering
  \subfigure[Block size 512]{
    \includegraphics[width=0.31\linewidth]{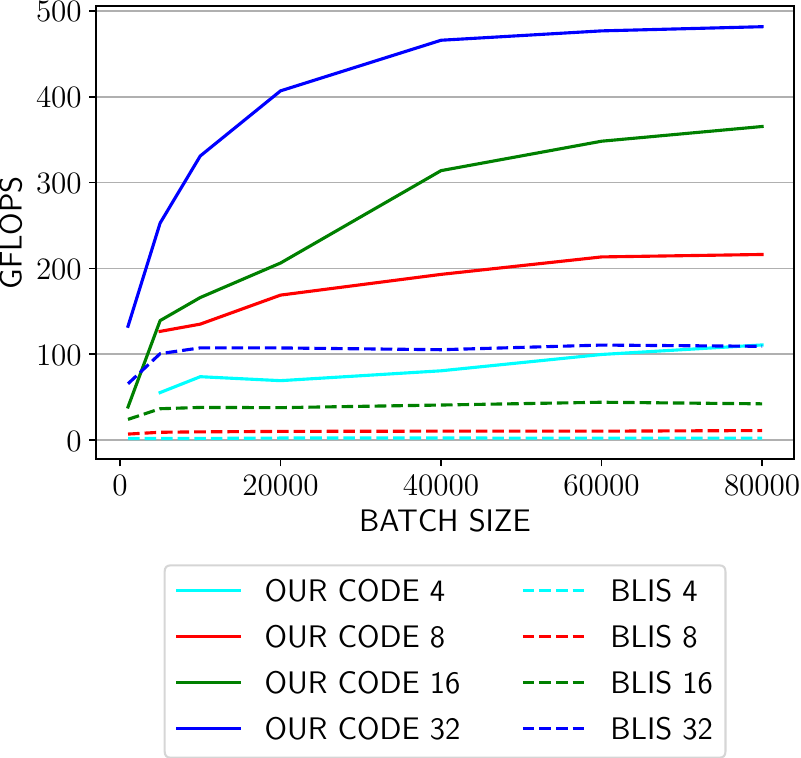}
    \label{fig:amd-threads-const-512}
  }
  \subfigure[Block size 1024]{
    \includegraphics[width=0.31\linewidth]{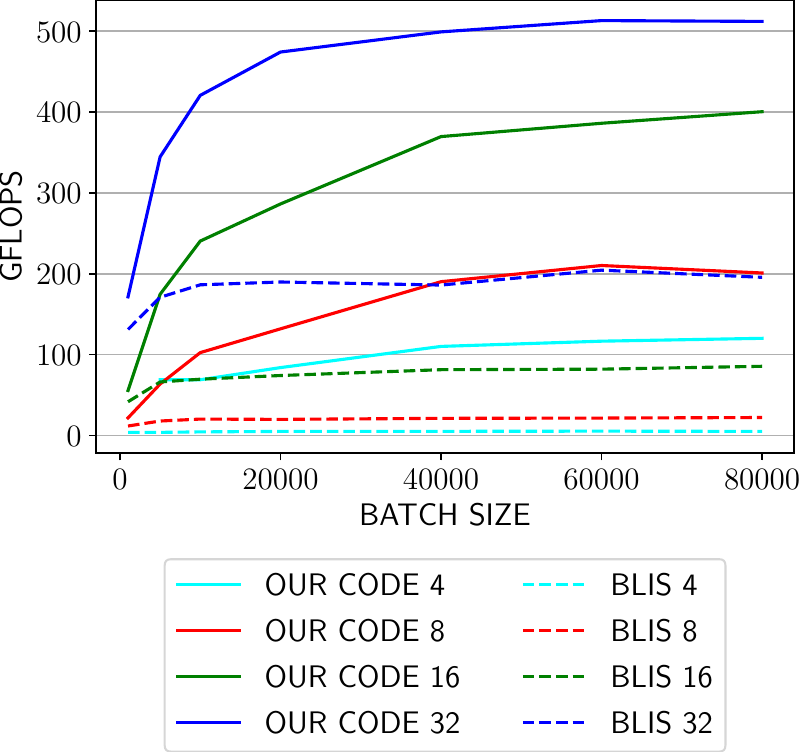}
    \label{fig:amd-threads-const-1024}
  }
  \subfigure[Block size 2048]{
    \includegraphics[width=0.31\linewidth]{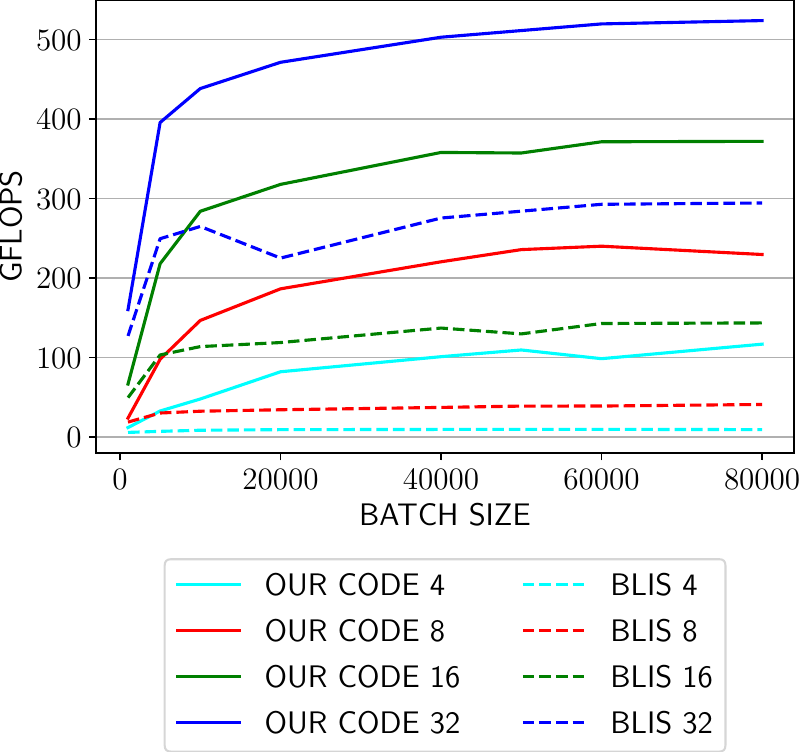}
    \label{fig:amd-threads-const-2048}
  }
  \caption{Performance for various block sizes and ranks when the 
    number of threads is constant at 64 for varying batch sizes for AMD EPYC 7502.}
  \label{fig:amd-threads-const-batch-variation}
\end{figure}

Fig. \ref{fig:amd-bandwidth-util} shows the bandwidth utilization of our code vs. AMD BLIS
for a variety of problem sizes. The bandwidth is calculated as shown in Eq. \ref{eq:amd-exp-bw-calculation}.
For ranks 4 and 8, the bandwidth increases in proportion to the performance for all block sizes, as shown
in Fig. \ref{fig:amd-gflops-batch-20k}. However, for rank 16 and 32,
we observe that the total utilized bandwidth does not saturate the available bandwidth
of the system, even though the corresponding GFLOPS utilization 
in \ref{fig:amd-gflops-batch-20k}
shows that the utilization for ranks 16 and 32 is much higher than that for 4 and 8. This
phenomena can be explained by the fact that the AMD EPYC CPU implements fully overlapping
caches as pointed out in Sec. \ref{sec:ecm-overlap-hypothesis}. Therefore, the computation
for rank 16 and 32 is more compute bound than memory bound when a 
sufficient number 
of physical cores are active. Compared to the bandwidth
behaviour of the Intel node in  Sec. \ref{sec:evaluation-skylake-x}, we can see
that even  though the AMD chip has
similar bandwidth, it is able to use the available memory bandwidth much more
efficiently as a result of the overlapping design of the caches.

\begin{figure}[tpb]
    \centering
    \includegraphics[width=\linewidth]{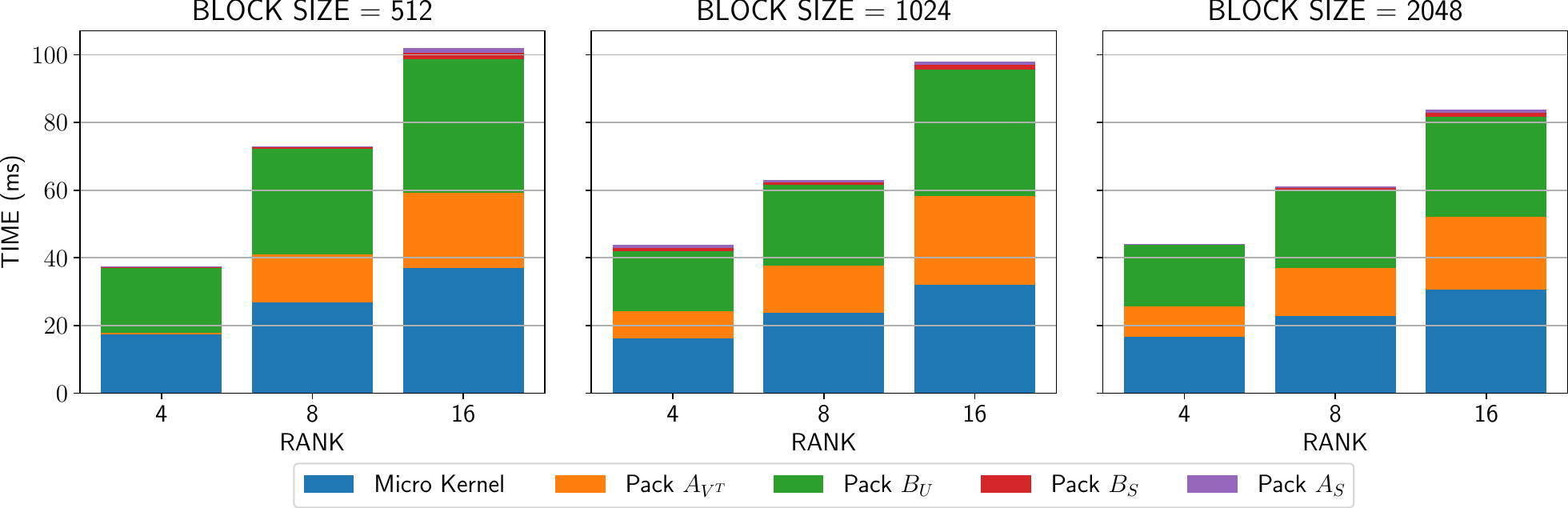}
    \caption{Performance breakdown for various parts of the computation for the 
        AMD node using 64 threads and a batch size of 20000.}
    \label{fig:perf-breakdown-amd}
\end{figure}

Fig. \ref{fig:amd-threads-const-batch-variation} shows the performance when
keeping the number of threads constant at 64 (i.e. using the full node) and changing
the batch sizes for variety of block sizes and ranks. While the AMD results differ
from the Intel results in the fact that the batch size has a non-trivial effect on
the utilization, our implementation still outperforms the AMD BLIS by a wide margin
for every problem size. Fig. \ref{fig:perf-breakdown-amd} shows the performance
breakdown for each part of the execution for the AMD node.

\rev{Table~\ref{tab:amd-higher-rank} shows the performance of our code vs. BLIS as the problem
gets more compute bound than memory bound. As shown for the Intel and Fujitsu nodes, the higher
performance of AMD-BLIS can be attributed to the fact that larger sizes of small matrices leads
to less optimal bandwidth utilization for our method.}

\begin{table}[]
\begin{tabular}{|l|l|l|l|l|}
\hline
\multicolumn{1}{|c|}{\textbf{Batch size}} & \multicolumn{1}{c|}{\textbf{Block size}} & \multicolumn{1}{c|}{\textbf{Rank}} & \multicolumn{1}{c|}{\textbf{BLIS}} & \multicolumn{1}{c|}{\textbf{OUR CODE}} \\ \hline
20000 & 512  & 96  &   197.056  & 237.441     \\ \hline
20000 & 1024 & 96  &   373.555  & 283.175    \\ \hline
20000  & 2048 & 96 & 559.075 & 348.566  \\ \hline
\end{tabular}
\caption{\rev{Performance of our code vs. BLIS for larger rank on the AMD node 
using 64 physical cores reported in GFLOPS. Our code can consistently beat AMD-BLIS
until rank 96 is reached.}}
\label{tab:amd-higher-rank}
\end{table}

\subsection{Evaluation of Block Low Rank matrix vector multiplication}
\label{sec:evaluate-blr-matvec}

\begin{figure}
    \centering
    \includegraphics[width=0.6\linewidth]{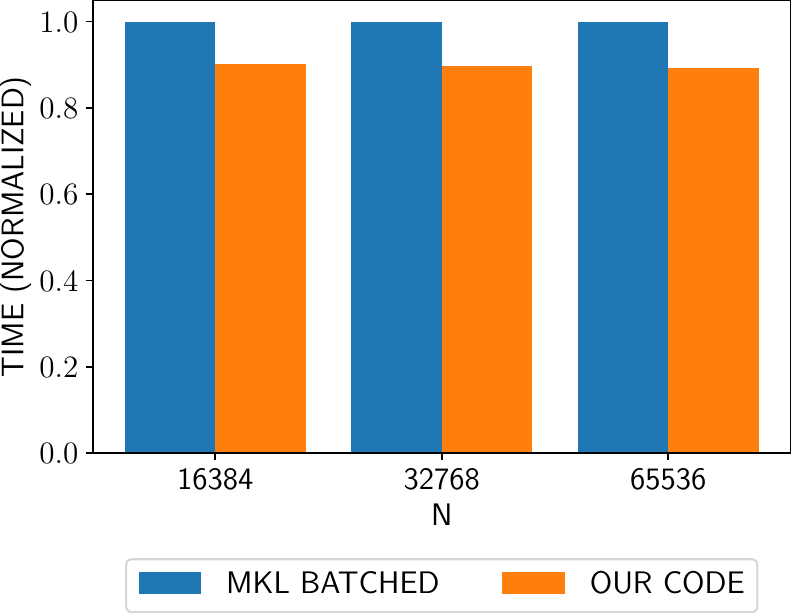}
    \caption{Multiplying multiple right hand sides with a weakly admissible
    block low rank matrix of various sizes using batched MKL vs. our code for the
    batched multiplication routine. The rank and number of right hand sides are
    kept constant at 8 for all tests.}
    \label{fig:evaluate-blr-matvec-abci}
\end{figure}

Fig. \ref{fig:evaluate-blr-matvec-abci} shows the comparison of run time for
multiplying multiple right hand sides using batched MKL vs. our implementation
of batched low rank multiplication on the Intel node. A gain of about 15\%
can be observed. While the multiplication of just the low rank
blocks and vectors shows a performance gain of about 50\%, adding the
intermediate vectors and multiplication of the dense blocks takes up more 
time, which reduces the performance gain to about 15\%.

Since we show in the previous sections that our batching methodology is about
twice as fast as vendor optimized libraries, we expect similar performance gains
for other nodes too.

%% file: conclusion.tex
\section{Conclusion and Future Work}
\label{sec:conclusion}

In this paper we have shown that performance of batched low rank multiplication 
can be improved with an alternative batching methodology based on improved 
data reuse and bandwidth utilization. 
Our results indicate better CPU utilization than vendor optimized libraries 
for a variety of thread counts and
batch sizes on 3 major CPUs architectures, and for problem sizes critical to low rank operations.
Specifically, we are able to achieve more than twice the performance of vendor
optimized matrix multiplication routines when using the entire node
for most problem cases. 
While most cases are memory bound, a larger rank of $32$ actually
results in a compute bound process. Thus our batching technique is able to keep
the SIMD units busy enough that bandwidth is no longer the bottleneck,
which is not true for vendor optimized libraries even for larger ranks.
We run the same algorithm on all the CPUs, thus the cache misses generated
by the data accesses would be proportional for the different libraries, thus 
the performance improvement comes from our optimization.

It is important to note that the constraints placed by the limited number of
registers in SIMD architectures places limitations on the scalability of 
our method.
However, hierarchical matrix factorization and vector multiplication 
typically involves low rank
multiplication with block size up to 2048 and batch sizes not exceeding 
20,000 -- ranges where our 
approach is highly competitive
with state of the art implementations, yielding significantly better results
for this specific problem.

In the future we plan to use our technique for building fast factorization
routines that run on distributed supercomputers. Distributed hierarchical
factorization is challenging due to irregularity of communication patterns and
computation. Cao et. al.~\cite{cao2019} report better load balance by using an
alternate process distribution and prioritization of the critical path using the
PaRSEC~\cite{bosilca2013}
runtime system. A more analytical approach~\cite{irony2004, solomonik2011}
leads to better understanding of the trade-off between replication and
communication of data for determining the data distribution on multiple nodes,
which can possibly lead to more efficient process distribution and
replication methodologies for minimization of communication overhead.